\documentclass[journal=langd5,manuscript=article]{achemso}

\usepackage{chemformula} 
\usepackage[T1]{fontenc} 
\usepackage[utf8]{inputenc}
\usepackage{lineno}

\usepackage{xr}

\author{Anne C. Nickel}
\affiliation[RWTH Aachen University]
{Institute of Physical Chemistry,  RWTH Aachen University, 52056 Aachen, Germany, European Union}
\author{Andrey A. Rudov}
\affiliation[DWI]
{DWI - Leibniz-Institute for Interactive Materials, 52056 Aachen, Germany, European Union}
\author{Igor I. Potemkin}
\email{potemkin@dwi.rwth-aachen.de}
\affiliation[DWI]
{DWI - Leibniz-Institute for Interactive Materials, 52056 Aachen, Germany, European Union}
\author{Jérôme J. Crassous}
\email{crassous@pc.rwth-aachen.de}
\affiliation[RWTH Aachen University]
{Institute of Physical Chemistry,  RWTH Aachen University, 52056 Aachen, Germany, European Union}
\author{Walter Richtering}
\email{richtering@rwth-aachen.de}
\affiliation[RWTH Aachen University]
{Institute of Physical Chemistry,  RWTH Aachen University, 52056 Aachen, Germany, European Union}

\title
 {Interfacial assembly of anisotropic core-shell and hollow microgels}

\begin{document}

\begin{abstract}
Microgels, cross-linked polymers with submicrometer size, are ideal soft model systems. While spherical microgels have been studied extensively, anisotropic microgels have been hardly investigated. In this study, we compare the interfacial deformation and assembly of anisotropic core-shell and hollow microgels. The core-shell microgel consists of an elliptical core of hematite covered with a thin silica layer and a thin shell made of poly(\textit{N}-isopropylacrylamide) (PNiPAM). The hollow microgels were obtained after a two-step etching procedure of the inorganic core. The behavior of these microgels at the oil-water interface was investigated in a Langmuir Blodgett trough combined with ex-situ atomic force microscopy. First, the influence of the architecture of anisotropic microgels on their spreading at the interface was investigated experimentally and by dissipative particle dynamic simulations. Hereby, the importance of the local shell thickness on the lateral and longitudinal interfacial deformation was highlighted as well as the differences between the core-shell and hollow architectures. The shape of the compression isotherms as well as the dimensions, ordering and orientation of the microgels at the different compressions were analysed. Due to their anisotropic shape and stiffness, both anisotropic microgels were found to exhibit significant capillary interactions with a preferential side-to-side assembly leading to stable microgel clusters at low interfacial coverage. Such capillary interactions were found to decrease in the case of the more deformable hollow anisotropic microgels. Consequently, anisotropic hollow microgels were found to distribute more evenly at high surface pressure compared to stiffer core-shell microgels. Our findings emphasize the complex interplay between the colloid design, anisotropy and softness on the interfacial assembly and the opportunities it thus offers to create more complex ordered interfaces.
\end{abstract}

\section{Introduction}
Colloids, particles in the size range of 1~nm to 1~µm dispersed in a continuous phase\cite{Butt2003}, are ideal model systems to investigate properties found in nature.\cite{Alsayed2005,Yunker2010,Dressaire2017,Mitragotri2008,Angelini2011,Parry2014} Soft anisotropically shaped objects like proteins, bacteria or complete cells contribute within many biological phenomena.\cite{Wosten1993,Park2003,Dell2018,Merkel2011,Vissers2018,Sullivan2019} Hence, anisotropic microgels can be ideal model systems to obtain more information about the influence of softness and anisotropy.\\
Microgels based on poly(\textit{N}-isopropylacrylamide) (PNiPAM) are highly interfacial active, lower the interfacial tension and are strongly deformed at oil-water interfaces. The interfacial phase behavior of such spherical microgels has been studied intensively showing the effect of size and hard cores to the spreading of spherical microgels at the interface and their response to an increase in surface pressure within the Langmuir-Pockels trough.\cite{Geisel2012,Mihut2013,Pinaud2014,Camerin2019,Rey2016,Picard2017,Bochenek2019,Scotti2019nc,Schmidt:2020kv,Ciarella:2021a,Harrer:2021,vialetto2021effect, Fernandez:2021}\\
Only few studies investigate the interfacial behavior of particles with an anisotropic shape.  Due to the specific shape, different ordering phenoma become relevant at the interface revealing tip-to-tip ($t-t$),\cite{Loudet2005} side-to-side ($s-s$)\cite{Loudet2005,Basavaraj2009,Pietra2012} or triangular\cite{Basavaraj2009} ordering for ellipsoidal particles. This ordering is a result of capillary interactions caused by a curved contact line to fulfill Young's equation.\cite{Lehle2008} Hence, the ordering of anisotropically shaped particles is influenced by multiple factors such as the particle shape and dimensions and the interfacial properties.\cite{Basavaraj2009,Botto2012}\\
The effect of softness onto the interfacial behavior of anisotropically shaped microgels was first introduced by Honda et al.\cite{Honda2019} and further developed in our previous study.\cite{Nickel2021} Hereby, the interfacial behavior of anisotropic core-shell microgels and more especially the influence of the microgel shell on their deformation, ordering and orientation at the interface were discussed.\cite{Nickel2021} This led to the conclusion that both the anisotropy of the hard core and the softness of the shell are key parameters defining the nearest neighbor ordering, the microgel orientation in respect to the barriers and the prominence of capillary interactions.\\
Many objects found in nature are soft and, hence, require a completely soft object as model. This reduces the applicability of core-shell microgels with a rigid core as optimal model system. Hence, a soft but still anisotropically shaped microgel is desirable. This can be achieved by removing  the hematite-silica core of anisotropically shaped core-shell microgels leading to hollow microgels.\cite{Nickel2019} Similar to spherical microgels\cite{Geisel2015}, the hollow design is expected to have a major influence on the spreading and assembly of anisotropic microgel at the interface. In order to investigate this effect, appropriate anisotropic hollow microgels that remain anisotropic at the interface must therefore be designed.\\
In this work, we study the influence of the rigid core on the behavior of anisotropic microgels at the fluid decane-water interface by comparing the behavior of a core-shell microgel with a thin shell to the resulting hollow microgel after etching the core. The shape and size of the microgels at the interface, and their collective and individual response to compression are investigated by combining Langmuir-Blodgett deposition and atomic force microscopy (AFM). Dissipative particle dynamics (DPD) simulations were performed to support the experiments and rationalize the interfacial deformation of the individual anisotropic microgels and the importance of the core-shell and hollow microgel design. We further analyzed and discussed the collective assembly and ordering of both systems at the interfaces from low to high compression.

\section{Experimental Section}
\textbf{Materials} All materials were used as purchased. The cross-linker BIS (\textit{N},\textit{N'}-methylene-bisacrylamide) was purchased from Sigma-Aldrich and the main monomer NIPAm (\textit{N}-isopropylacrylamide) from Acro organics. The initiator KPS (potassium peroxydisulfate) was bought from Merck. The co-monomer BAC (\textit{N},\textit{N}’-bis(acryloyl)cystamine) was purchased from Alfa Aesar. For the Langmuir-Blodgett trough experiments decane (Merck) , ultrapure water (Astacus$^{2}$, membraPure GmbH, Germany) and propan-2-ol (Merck) were used. The decane was cleaned by filtering it three times over a column of basic Al$_{2}$O$_{3}$ (Merck). The third filtering was done just before the measurement. \\
\textbf{Synthesis} The synthesis of the core-shell microgel ($CS-165$) is described in detail in a former study\cite{Nickel2021}. Briefly, 
584.4~mg of NIPAm, 39.6~mg of BIS and 14.3~mg of BAC were dissolved in 280~mL of filtered bidistilled water and 3~mL of an ethanol solution with 1.4~wt\% of ellipsoidal functionalized hematite-silica cores were added into the three-neck flask after sonification for 30~minutes. The mixture was heated with an external oil-bath to 60~°C and degased with nitrogen. After 45~Minutes, the oil-bath was set to 80~°C and after another 15~Minutes, the polymerisation was started by adding the initiator solution of 41.0~mg KPS degased in 10~mL filtered bidestilled water.  dropwise. The reaction was stopped after 4~h and the microgels were purified with three times centrifugation at 5000~rpm for 60~Minutes and redispersing in filtered bidestilled water. The hollow microgels ($HS-165$) are obtained after the two-step etching of the $CS-165$. The hematite was first etched under acidic conditions using concentrated hydrochloric acid via dialysis. Afterwards the solvent was exchanged with ultrapure water until the pH was neutral again. The remaining silica shell was etched under alkaline conditions using dialysis against 0.5 molar sodium hydroxide. The solvent was exchanged to obtain a neutral pH after the silica was etched away and the resulting microgels were lyophilizates.\\
\textbf{Characterization}\\
A detailed characterization of the microgels is described in former studies\cite{Nickel2019,Nickel2021}. The most important microgel characteristics are summarized in Table~\ref{tab_microgels}.
\begin{table}[H]
     \centering
    \caption{Characterisation of the synthesized microgels.}
     \begin{tabular}{ccc}
     \hline
     & $CS-165$ & $HS-165$\\
       \hline
       dimensions of the core & (330 $\pm$ 12) nm $\times$ (75 $\pm$ 8) nm & - \\
       polymer content & 65.0\% & 100\% \\
      $ R_{H}^{10~^{\circ}C}$ & (165 $\pm$ 2) nm & (182 $\pm$ 2) nm \\
      $  R_{H}^{50~^{\circ}C} $ & (122 $\pm$ 1) nm & (127 $\pm$ 1) nm\\
     \end{tabular}
     \label{tab_microgels}
     \end{table}

\noindent \textbf{Langmuir-Blodgett technique.} The isotherms and depositions were conducted simultaneously at a decane-water interface in a customized liquid-liquid Langmuir-Pockels trough (KSV NIMA, Biolin Scientific Oy, Finland) equipped with two movable barriers. The exact procedure is described in Ref.~\cite{Nickel2021}. The trough was cleaned with ethanol and milli-q water before a clean water-decane interface was established. To this interface, the microgel dispersion to which isopropanol was mixed as spreading agent were added with a syringe. The substrates were cleaned before they were moved through the water-decane interface. The microstructures of the microgels on the substrates were further investigated with atomic force microscopy (AFM).\\
For the hollow microgel $HS-165$, it was not possible to obtain the full compression isotherm within one single measurement with the used Langmuir-Blodgett trough. Hence, a compression isotherm had to be measured with an initial surface pressure ($\Pi$) larger than zero. The values for the compression isotherm obtained at the highest $HS-165$ concentration were rescaled by 5~mN/m each to overlap all three compression isotherms conducted on this system. The original data can be found in the supporting information (Figure~S4).\\

\noindent \textbf{Atomic force microscopy.} Dry-state AFM was performed to visualize the monolayers deposited on the silica substrates. The experiments were carried out in tapping mode using a Dimension Icon AFM with a closed loop (Veeco Instruments Inc., USA, Software: NanoScope 9.4, Bruker Co., USA). The tips used for imaging were OTESPA tips with a resonance frequency of 300~kHz, a nominal spring constant of 26~N$\cdot$m$^{-1}$ of the cantilever and a nominal tip radius of <7~nm (NanoAndMore GmbH, Germany). The images were recorded with the programmed move function to scan the silica substrates. On each substrate, six rows along the gradient of microgel concentration were captured with a step-width of 1000~µm. Each image is 7.5~µm~$\times$~7.5~µm large with a resolution of 512~pixels~$\times$~512~pixels. Higher resolution images were obtained by using the same number of pixels (512~pixels~$\times$~512~pixels) for an image size of 3~µm~$\times$~3~µm. Such micrographs were captured additional to the programmed move.\\

\noindent \textbf{Image Analysis} The program NanoScope Analysis 1.9 was used for the acquisition and leveling of the AFM micrographs. The processed AFM images were analysed with an updated version of the MATLAB routine of Bochenek et al.\cite{Bochenek2019,Nickel2021}, details are provided in the supporting information. \\

\noindent \textbf{Computer simulations} \\
\noindent \textit{Method.} To assess the difference between the behavior of anisotropic core-shell (CS) and hollow-shell (HS) microgels at the decane-water interface DPD simulations were used\cite{Espanol2017}. The method was successfully applied and described in detail for the study of anisotropic hybrid core-shell microgels with a different shell thickness\cite{Nickel2021}. \\The modeling of the microgels, which have a size of hundreds of nanometers, requires a proper coarse-graining procedure\cite{Spaeth2011} leading to the effective reduction of the gel size. Dealing with the hollow anisotropic microgels, we found it crucial to readjust a coarse-grained relation between the size, softness of the microgel, and surface/volume microgel-solvent interactions. The higher the coarse-graining degree, the higher the sensitivity of the system to the disbalance between surface and volume terms, which is especially noticeable for oblong or hollow objects at the same coarse-grained level. Thus the size-limit effect (see Supporting information, Size-limit effect) must be considered. In the case of regular or hollow spherical microgels the effect is less visible due to the spherical symmetry of the system. The swelling/deswelling process is accompanied by the self-similar change in the size of the microgel and its cavity. Both, in collapsed and swollen states, the hollow microgel and its cavity are spherically shaped \cite{Geisel2015, Brugnoni2018, Brugnoni2020}. This result is reproduced in a fairly wide range of interaction parameters as well as the flexibility of the subchains\cite{Song2019, Scotti2018}. However, following such a strategy one could notice, that simulation of the hollow ellipsoidal microgels failed and could not reproduce the experimental observations\cite{Nickel2019}. In particular, different trends of the shape/size evolution of the shell and the void upon collapse above their volume phase transition temperature (VPTT) are observed or extra flattening of the microgel at the interfaces is detected. \\
In this study, we extended the method used earlier and applied several modifications to improve the modeling of the  CS and HS anisotropic microgels in bulk and at the decane-water interface. The first modification is aimed to choose the reasonable coarse-grained level and reproduce interfacial tension value for the pure decane/water interface without the microgel. Decane molecule is treated as a single bead in the system similarly to Ref.\cite{Alasiri2017}. The DPD interaction parameters $a_{ij}$ were estimated through the Flory-Huggins interaction parameter matching the infinite dilution activity coefficient of liquids, where $i$ and ${j \in }$(W - water bead, D - decane bead) see Supporting information, Decane/water interface. Such a choice of parameters at 20 $^\circ C$ provides the value of the interfacial tension of water/decane liquids equal to 51.9 mN/m which is in good agreement with the experimental value 52.33 mN/m.\cite{Zeppieri2001}\\ 
The second modification is related to the model of the $CS-165$ and $HS-165$ microgels. The anisotropic CS microgel is composed of a polymeric shell and a solid core. The beads forming the core and the shell are denoted by $C$ and $S$, respectively.
 The dimensions of the hematite-silica core in the experiment were 330$\pm$12nm$\times$75$\pm$8 nm. The ellipsoidal nanoparticle having similar aspect ratio between the long and short semi-axes 4.4$\times$1 have been created. Since the real aspect ratio of the CS-165 is unknown, we designed three different microgels with the same shell mass and cross-linking density but different initial aspect ratios of the shells $L/d$ = 2.8; 2.5 and 2.2 to capture the peculiarities of the shell dimensions (Figure S5, Table S3). The microgel with a shell aspect ratio of 2.8 is featured by a thicker shell at the top of the core compared to its side. The microgel with an aspect ratio of 2.2 has approximately the same thickness of the shell around the solid particle. Hollow microgels were obtained after removal of the solid core from the corresponding core-shell microgels (see Supporting information, Microgel description). \\
In our previous work \cite{Nickel2021}, we assumed that independently of the shell thickness $CS$-samples has a 5 mol\% cross-linker groups homogeneously distributed within the gels. Indeed, the molar mass composition of the cross-linker BIS in experiments was around 4.69, 4.63 and 4.72 mol\% for $CS-356$, $CS-254$ and $CS-165$, respectively. However, related to the specificity of the synthesis of $CS-165$, the exchange of the dye methacryloxyethyl thiocarbamoyl rhodamine B (MRB) to BAC both introduce additional cross-links and may change the polymerization kinetics resulting in a stiffer microgel shell and a lower degree of swelling. The DLS measurements and results of the mass loss during the thermogravimetric analysis for the CS-254 and CS-165 microgels with different shell thicknesses confirmed our conclusions. The samples have a similar amount of shell related to the core mass: 60.9\% and  65.0\% respectively which explain similar hydrodynamic radii in the collapsed state\cite{Nickel2021}. At the same time in the swollen state at 20 °C, $CS-254$ is around 90 nm large than $CS-165$ (the swelling degree of CS-254 and CS-165 is equal to 2.2 and 1.4, respectively) which indicates a higher cross-linking density of $CS-165$ sample. We found that in simulation an approximate degree of crosslinking, which could reproduce the swelling degree of the $CS-165$ microgel in water, is equal to 7.5\%. The average subchain length, $N$, equals $6\pm1$ (see Supporting information, Microgel characterization in bulk).\\
 The third modification provides the way to reproduce the swelling degree of the microgel at chosen coarse-grained level by establishing the relation between the number of the beads per subchain and the subchain flexibility. Similarly to the work by Nikolov et al.\cite{Nikolov2018} in addition to the harmonic potential, a bending angle potential as well as a segmented-repulsive potential (SRP)\cite{Sirk2012} between the polymer beads have been introduced. The SRP potential prevents polymeric subchains from crossing one another. The bending potential enables to control the Kuhn length of the subchains and thus the flexibility of the subchains and the entire microgel. Based on the idea that in a swollen state PNiPAM polymers exhibit flexible behavior, we determined bending stiffness value for the subchain length N=6 (see Supporting information, Microgel description). In order to reproduce the swelling/deswelling behaviour of the microgels in the liquids, we alter the polymer-liquid repulsion ${a_{Si}}$ in a range between $a_{Si} = 25 k_BT/r_c$ and $a_{Si} = 30.2k_BT/r_c$, where $i$ ${ \in (W,O) }$. Following that strategy we found suitable parameters to reproduce both the experimental swelling degree of the microgels in the bulk and their deformability at the interface.\\
Detailed information about simulation parameters, modifications and its impact are fully described in Supporting information.

Simulations were performed using a DPD code supplied by Lammps\cite{Plimpton1995}. Simulations of microgel in a solution were performed in the $NVT$ ensemble. Cubic boxes of a constant volume $V = L_{x}$ $\times$ $L_{y}$ $\times$ $L_{z}$ = 100 $\times$ 100 $\times$ 100 $r_{c}^3$ with imposed three-dimensional periodic boundary conditions, containing single microgel immersed into the solvent were simulated for an equilibration phase of $10^6$ steps followed by a production phase comprising of $10^6$ steps. Equilibrium was deemed to have been achieved when the cumulative average temperature, pressure, gyration radii reached a plateau. 
The simulations of microgel at water/decane interface were performed in the $NVT$ ensemble in rectangular box of a constant volume $V = L_{x}$ $\times$ $L_{y}$ $\times$ $L_{z}$ = 140 $\times$ 140 $\times$ 37 $r_{c}^3$ with periodic boundary conditions in all directions. The box were fulfilled by water and decane beads. Water and decane phase-separate. Water is at the bottom of the simulation box, decane at the top. Note that the $z$ direction of the simulation box is perpendicular to the liquid/liquid interface. The interfacial tension between the two liquids is determined using the pressure tensors as shown in Eq. \ref{simeq9}. The microgel was placed close to the water/decane interface. The equilibration run comprised $10^6$ steps followed by a production phase comprising of $10^6$ steps. The interfacial tension, the gyration tensor and the surface area of the microgel, were averaged over the last $10^6$ steps of a total $2\cdot 10^6$ steps.\\

\noindent \textbf{Computational details}\\
\noindent \textit{Interfacial tension (IFT).} The DPD IFT (${\gamma_{DPD}}$) was evaluated by dividing the simulation box into a number of slabs parallel to the interface and obtaining the ensemble average of pressure tensors which can be presented by the following equation\cite{Maiti2004}:
\begin{equation}
    IFT_{DPD} = 0.5\int [<p_{zz}(z)>-\frac{1}{2}(<p_{xx}(z)>+<p_{yy}(z))>]\,dz 
        \label{simeq9}
\end{equation}
\noindent where rescaled $z = z/r_c$ and the angle brackets indicate the ensemble average of local dimensionless tensors components ($p_{xx}$, $p_{yy}$ and $p_{zz}$) of $x$, $y$ and $z$ directions. In application of Eq. \ref{simeq9}, the interface between two immiscible fluids is considered as a plane perpendicular to the $z$-direction. The evaluated ${\gamma_{DPD}}$ which is in DPD units and can be converted into conventional or physical IFT which is measured experimentally, by using the conversion factor \cite{Maiti2004} of $k_BT/r_c^2$  (i.e., $\gamma[mN/m] = ( k_BT/r_c^2 )\cdot\gamma_{DPD})$.\\

\noindent \textit{Gyration tensor.} The gyration tensor was calculated using the followings equation:
\begin{equation}
    \boldsymbol{S} = \frac{1}{N} \sum\nolimits_{i = 1}^{N} \boldsymbol{s}_{i}\boldsymbol{s}_{i}^{T} = \overline{\boldsymbol{ss}^{T}} = \left[ \begin{array}{rrr}
\overline{x^{2}} & \overline{xy} & \overline{xz}\\ 
\overline{yx} & \overline{y^{2}} & \overline{yz} \\
\overline{zx} & \overline{zy} & \overline{z^{2}} \\ 
\end{array}\right],
    	\label{simeq10}
\end{equation}
where $\boldsymbol{s}_{i} = \left[ \begin{array}{r}
x_{i} \\ 
y_{i} \\
z_{i}\\ 
\end{array}\right] $ is the position vector of each bead, which is considered with respect to the center of mass of the microgel $\sum\nolimits_{i = 1}^{N} \boldsymbol{s}_{i} = 0$ , and the overbars denote an average over all the beads in the microgel, $N$. To get a complete characterization of the elliptic microgels, we computed the eigenvalues of the gyration tensor $\lambda_{x} = \overline{X^{2}}$, $\lambda_{y} = \overline{Y^{2}}$, $\lambda_{z} = \overline{Z^{2}}$ as well as eigenvectors $\overline{T_{1}}$,$\overline{T_{2}}$,$\overline{T_{3}}$. The gyration tensor is symmetric thus, the Cartesian coordinate system can be found in which it is diagonal. Transformation to the principal axis system diagonalizes $S$, and we chose the principal axis system in which

\begin{equation}
    \boldsymbol{S} = diag(\lambda_{x},\lambda_{y},\lambda_{z}),
    	\label{simeq11}
\end{equation}
\noindent where we assume that the eigenvalues of $\boldsymbol{S}$ are sorted in ascending order, i.e., $\lambda_{x}\leq\lambda_{y}\leq\lambda_{z}$ and $x$, $y$ and $z$ are the new coordinate axes. These eigenvalues are called the principal moments of the gyration tensor. The first invariant of $S$ gives the squared radius of gyration,

\begin{equation}
    tr(\boldsymbol{S}) \equiv I_{1} = \lambda_{x} + \lambda_{y} + \lambda_{z} = R_{g}^{2},
    	\label{simeq12}
\end{equation}

\noindent a measure of the average size of the particular conformation of the microgel. We also defined the shape anisotropy of the microgel as
\begin{equation}
    L/d = \lambda_z/\lambda_{x,y}
    	\label{simeq13}
\end{equation}
\\
\noindent \textit{Form factor}. The form factor P(q) of microgel in bulk was calculated using the followings equation:
\begin{equation}
    P(q) = \frac{1}{N} \sum_{ij} <e^{(-i\boldsymbol{(q r_{ij})}})>
    	\label{simeq14}
\end{equation}
\noindent where the angular brackets indicate an average over different equilibrium configurations of the same microgel and over different orientations of the wavevector $\boldsymbol{q}$, $\boldsymbol{r}_{ij}$ is the radius vector between the $i^{th}$ and $j^{th}$ beads. The fitting of the form factors were computed with the program sasview v.5.0.4 with a core-shell ellipsoid model.

\section{Results and Discussion}

\textbf{Compression isotherms and interfacial deposition.}
In this study, a hybrid core-shell microgel ($CS-165$) and its corresponding hollow counterpart ($HS-165$), obtained after its core dissolution, were employed to investigated the influence of microgel architecture on the interfacial deformation and assembly of anisotropic microgels. Core-shell microgels with larger and softer shells introduced in previous studies\cite{Nickel2019,Nickel2021}, did not exhibit any anisotropic shell at the interface. Conversely, the microgel $CS-165$ with the thinnest and stiffest shell exhibited an anisotropic shape at the interface\cite{Nickel2021} and was therefore selected for this study to investigate the differences between the behavior of anisotropic core-shell and hollow microgels at a decane-water interface.

\begin{figure}[ht]
\centering
\includegraphics[width=1\linewidth, trim={0cm 0cm 0cm 0cm},clip]{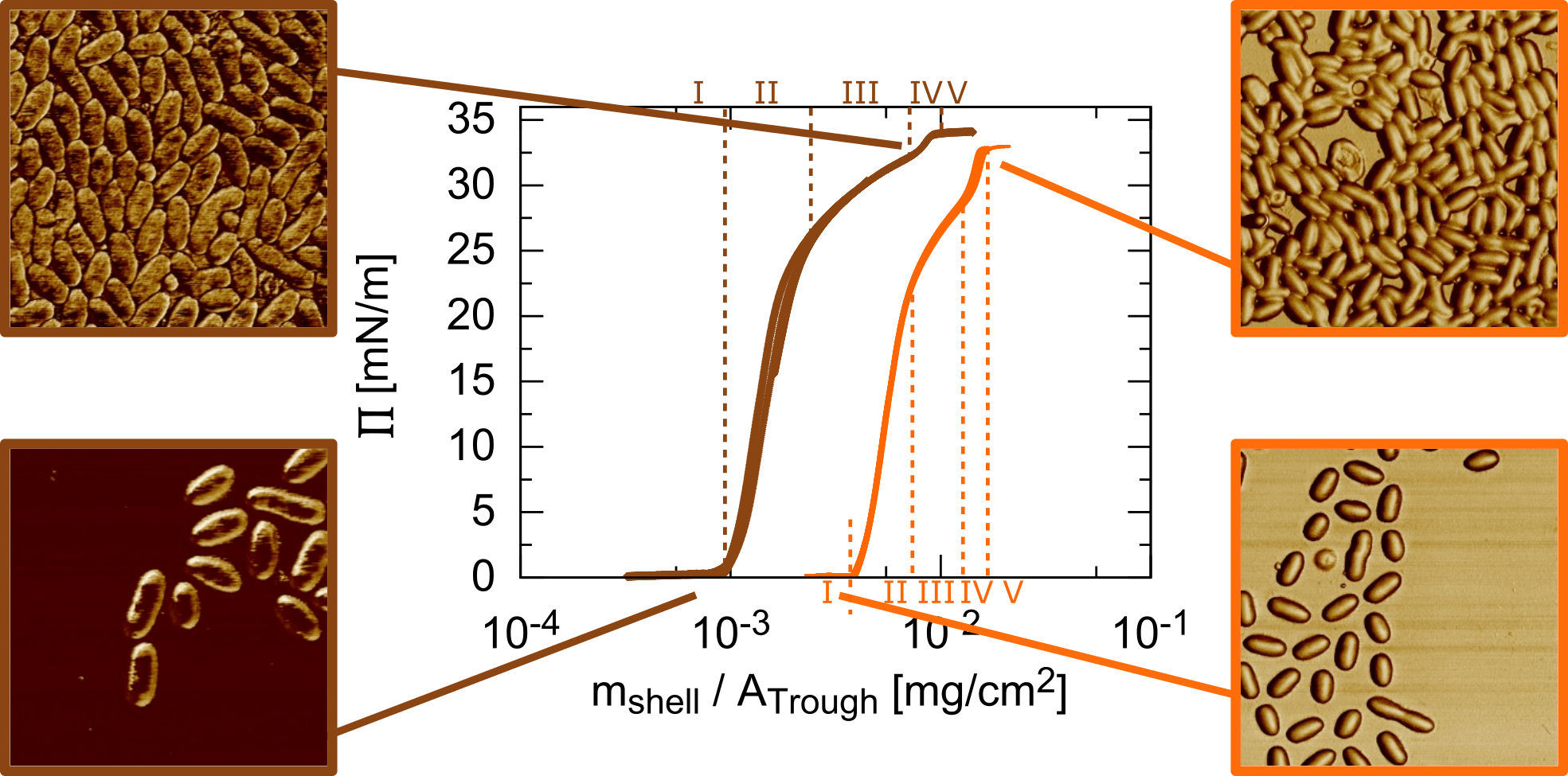}
	\caption{AFM phase images of hollow (left, brown) and core-shell (right, orange) microgels in a concentrated region (upper row) and a diluted region (lower row). Center: Compression isotherms of hollow (brown) and core-shell (orange) microgels in dependence of the mass of the shell divided by the trough area. Both compression isotherms reveal five regions. The AFM images have a size of 3~µm~$\times$~3~µm. The images and data of the core-shell microgels are reproduced from Nickel et al. Anisotropic Microgels show their Soft Side. Langmuir 2021, https://doi.org/10.1021/acs.langmuir.1c01748 Copyright 2021 American Chemical Society}
	\label{HCS_Isothermen}
\end{figure}

The microgels were investigated at the water/decane interface using a Langmuir-Blodgett setup. Subsequently, the dried deposited monolayer was analyzed via atomic force microscopy (AFM) to obtain microstructural information at the different compression states. The compression isotherms of hollow ($HS-165$, brown) and core-shell ($CS-165$, orange) microgels in dependence of the mass of the shell per area of the trough are shown in Figure~\ref{HCS_Isothermen}. The AFM phase micrographs in Fig.~\ref{HCS_Isothermen} evidence the anisotropic shape of hollow (left) and core-shell (right) microgels, respectively, in the concentrated (upper row) and diluted (lower row) regions. The AFM images have a size of 3~µm~$\times$~3~µm and demonstrate that even in the concentrated state, the microgels appear elongated independently of the presence or absence of the anisotropic core. Hence, the two microgel systems are suitable to identify the effects of the internal architecture on the interfacial behavior of anisotropic microgels. Furthermore, the hard core of core-shell microgels makes deformations such as bending impossible and the elliptical shape is preserved also at high surface pressures. The hollow microgels without the core obtain the freedom to slightly deform which can be observed at high surface coverage. This means they still have an anisotropic shape but can bend to better fit into observed microgel clusters. The microgels form clusters which do not show a direct contact in the dilute regime. This could be explained by either a polymer layer which is too thin to be visualized with dry-state AFM or by a deswelling of the outer polymer layer due to the drying at the silicate substrate.

$\Pi$ is reported as function of mass of the shell $m_{shell}$ normalized by the trough area, $A_{Trough}$. Considering that both microgels have the same microgel shell, the compression isotherms are therefore compared in respect to the number of microgels adsorbed at the interface. This representation of the data was selected, as the clustering observed for both microgels did not allow for an accurate normalization by the number of microgels observed within the AFM images. For $HS-165$, the total mass of the sample was considered as the microgel consists of the polymeric shell only, whereas for $CS-165$, 65~\% of the deposited mass corresponds to the polymeric shell as determined with thermogravitometry (TGA) in our former study.\cite{Nickel2021}

Both compression isotherms exhibit five distinct regions with increasing surface pressure (Figure~\ref{HCS_Isothermen}~center). Former studies on spherical microgels reported a similar two-step increase in surface pressure, where the first increase was attributed to the interactions of the uncompressed adsorbed microgel coronas and the second to the interactions between the harder cores.\cite{Geisel2015,Rey2016,Bochenek2019,ciarella2021soft} For hard ellipsoidal particles, compression isotherms differ with their aspect ratio.\cite{Anjali2019} While spherical hard particles and particles with an aspect ratio of 9 showed only one increase in surface pressure,\cite{Anjali2019} intermediate aspect ratios of 5.5 exhibited a two-step increase.\cite{Basavaraj2006} A flipping of the ellipsoidal particles at high surface pressure causes a decrease in the occupied area and that leads to the formation of the intermediate plateau in the compression isotherm.\cite{Basavaraj2006}\\

For the hollow and core-shell microgels, no flipping was expected due to the spreading of the microgel shells. The two-step increases of the surface pressure in the compression isotherms observed for both microgels are more similar to the two-step increase observed for spherical microgels as compared to those reported for hard ellipsoids.

Region I in Fig.~\ref{HCS_Isothermen} is characterized by a constant, zero $\Pi$ indicating that the concentration of the microgels is too low to influence $\Pi$. In region I\hspace{0.05cm}I, a sharp increase in surface pressure is observed for both microgels. Region III is ended by a sharp increase in surface pressure (region IV) followed by a constant value for the surface pressure (region V). Even if both microgels show the same number of regions, the shape of their isotherms differs. The main difference lies in the width of region III and the steepness of the surface pressure increase in region IV. The hollow microgel has a more extended region III and only a small increase in surface pressure in region IV. In comparison, the core-shell microgels has a much narrower region III and presents a steeper increase in surface pressure in region IV. For spherical microgels, an enlargement of region III is related to the softness of the microgels that can be either achieved for core-shell microgels with a hard silica core by decreasing the microgel shell crosslinking or by the removing the core.\cite{Geisel2015, vialetto2021effect}. Similarly for the anisotropic microgels investigated in this study, $HS-165$ appears softer than $CS-165$ leading to an enhanced spreading at the interface and an enlargement of region III.\\

Geisel et al.\cite{Geisel2015} investigated the compression modulus ($C^{-1} = -A(d\Pi/dA)$) for spherical hard core-shell and the corresponding hollow microgels. Two maxima were observed for the compression modulus revealing a two step-increase within the compression isotherm. For spherical microgels, the existence of the hard core did not influence the height of the first maxima, but rather the second maximum.\cite{Geisel2015} The same accounts for the anisotropically shaped microgels in this study (see SI for further information). The dense anisotropic microgel monolayers show a comparable elasticity as the one reported for spherical microgels.\\
This is inferred by comparing the data of the anisotropic microgels (Figure~S1) with the data obtained for spherical core-shell microgels from Geisel et al.\cite{Geisel2015}. Spherical microgels with a cross-linking density of 5~mol\% have the first maximum at $\Pi$ = 10~mN/m for $C^{-1}$= 46-47~mN/m~\cite{Geisel2015}. Although AFM phase images and dynamic light scattering data of the anisotropic microgels\cite{Nickel2021} pointed to a stiff microgel network, the compression modulus indicates a similar elasticity at the interface as was found for spherical microgels with similar cross-linker content.\cite{Geisel2015}\\

\begin{figure}[ht].
\centering
\includegraphics[width=0.6\linewidth, trim={0cm 0cm 0cm 0cm},clip]{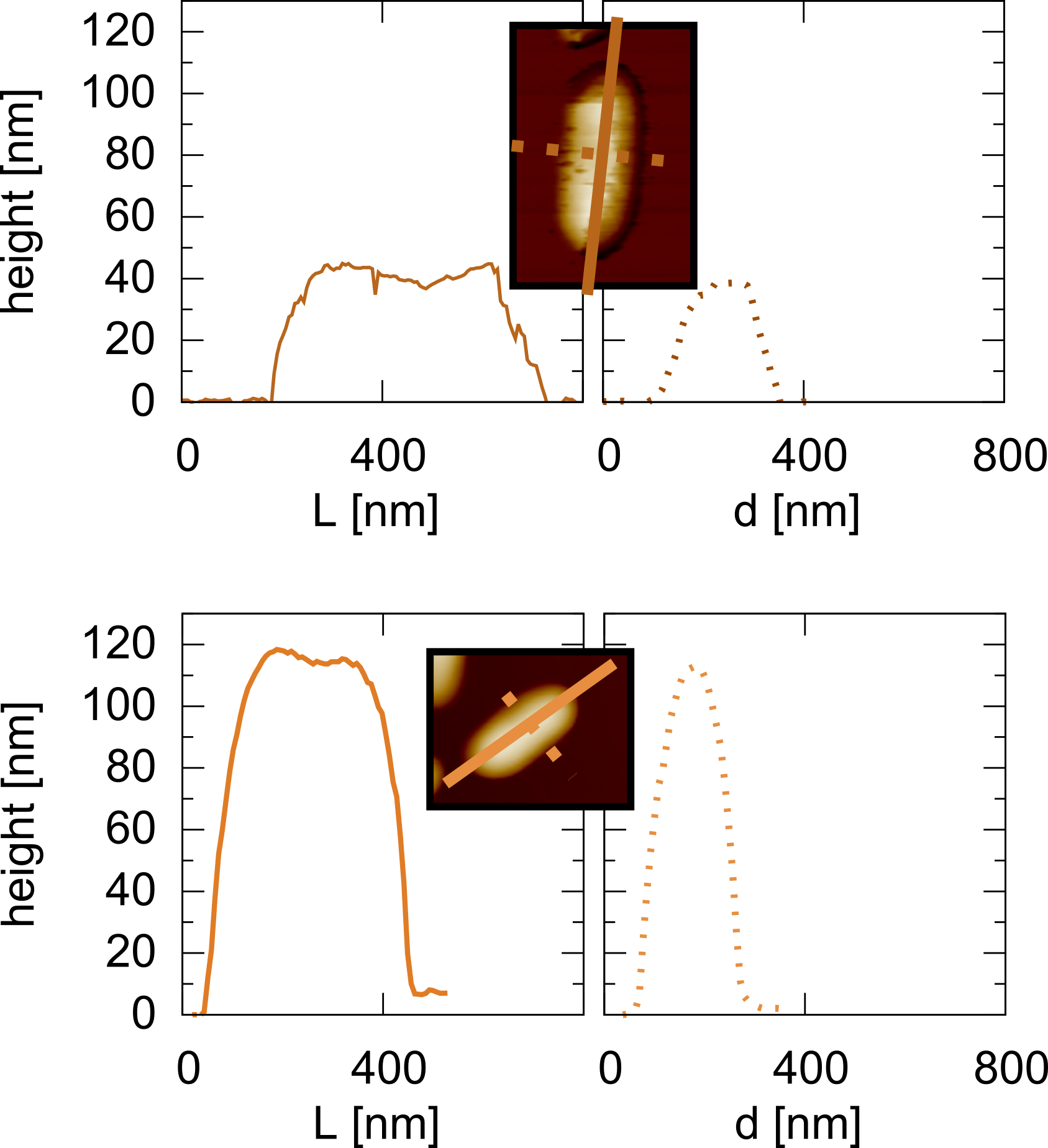}
	\caption{Height profiles of the long and short axes of one microgel of $HS-165$ (top) and $CS-165$ (bottom) obtained from AFM height micrographs.} 
	\label{HCS_height}
\end{figure}

The onset of the surface pressure increase takes place at lower compression for the hollow as compared to the core-shell microgels. Geisel et al.\cite{Geisel2015} observed a similar shift when comparing the compression isotherms of spherical core-shell and hollow microgels. This can be explained by the different size of hollow and core-shell microgels at the interface. Height profiles and the determination of the long and short semi-axes are shown in Figure~\ref{HCS_height} The mean area of one microgel increases from 0.0607$\pm$0.0055~µm$^{2}$ for $CS-165$ to 0.127$\pm$0.022~µm$^{2}$ for $HS-165$. By removing the core, the microgel shell can spread freely at the interface and the area covered by one microgel roughly doubles in size. This increase in area comes from an increase in the microgel length and diameter while the aspect ratio of the microgels stays nearly constant (see Table~\ref{tab_lengthwidth}). The length of the microgels increases by 283~nm from $CS-165$ to $HS-165$ while the additional spreading at the sides is 160~nm from $CS-165$ to $HS-165$.

Fig.~\ref{HCS_height}~(top) illustrates an example of the rapid decrease and increase of the height profile along the long and short axes of one ellipsoidal microgel. A smooth profile with a nearly constant height is observed for $CS-165$. For $HS-165$, the elliptical core was etched away, which leads to a rougher and flattened height profile and a much lower overall height (Fig.~\ref{HCS_height}~bottom).  \\

\begin{figure}[H]
\centering
\includegraphics[width=0.9\linewidth, trim={0cm 0cm 0cm 0cm},clip]{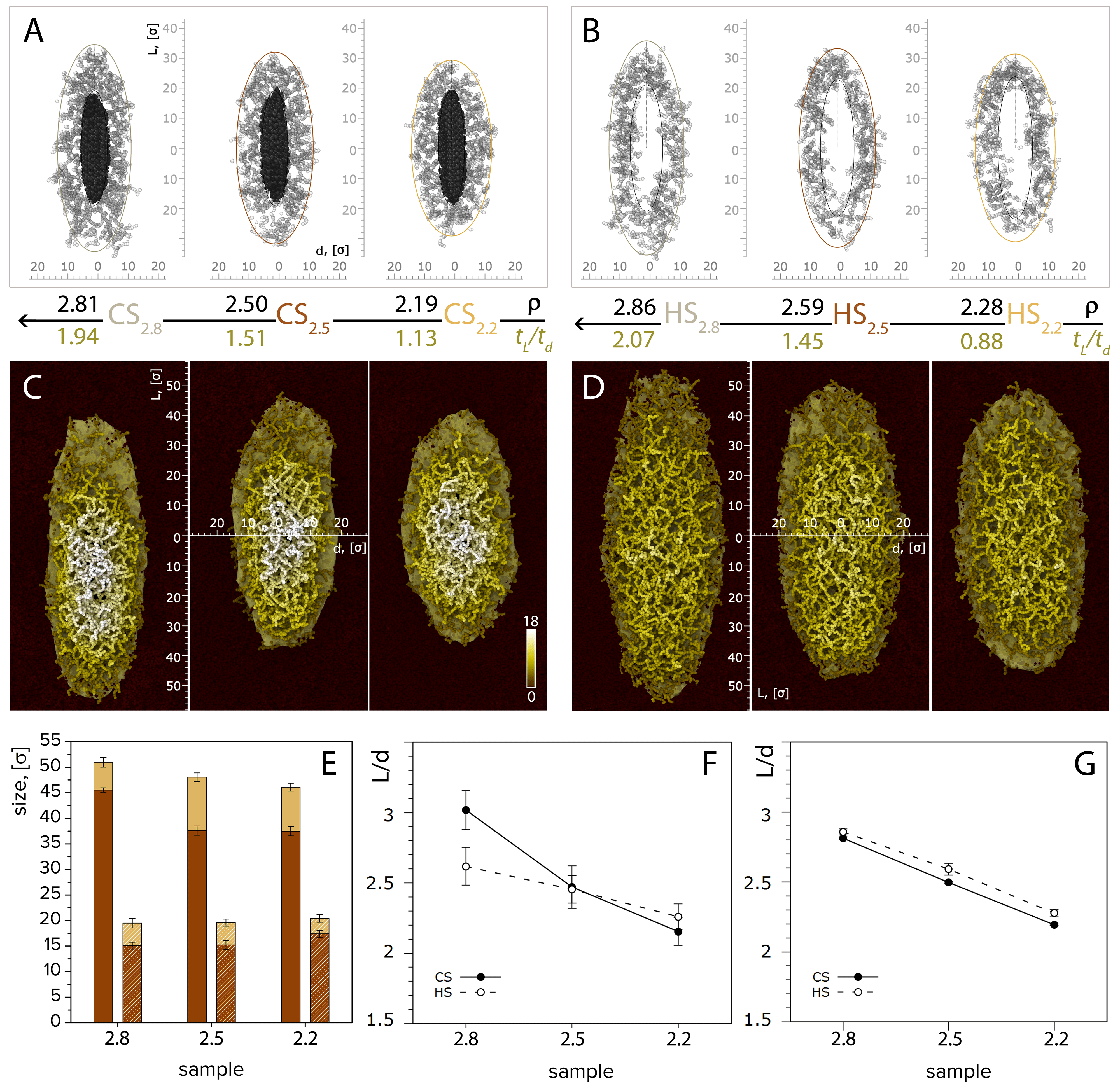}
	\caption{Top row: Equilibrium structure of the swollen $CS$ (A) and $HS$ (B) microgels with different initial shell configuration: $L_{frame}/d_{frame} = 2.8, 2.5$ or $2.2$ together with their corresponding overall aspect ratio, $\rho$ and the ratio of the longitudinal and lateral shell thickness, $t_L/t_d$. Middle row: Color height maps of the core-shell, $CS_{2.8}, CS_{2.5}, CS_{2.2}$ (C) and the hollow, $HS_{2.8}, HS_{2.5}, HS_{2.2}$ (D) anisotropic microgels at the decane/water interface. Top view looking from the water phase. (E) The values of the major, $L$, and minor, $d$, axes of the hollow (orange) and the core-shell (brown) anisotropic microgel obtained by the fitting of the contact area of the mirogels at the water/oil interface. Interfacial (F) and bulk (G) aspect ratio, $\rho_{int}$, for different samples. Solid and dashed lines correspond to $CS$  and $HS$ anisotropic microgels, respectively.} 
	\label{sim_1}
\end{figure}

\noindent \textbf{Interfacial deformation of individual anisotropic microgels: comparison with simulations.}
The conclusions about shape and spreading of individual particles in the uncompressed region I are nicely supported by the computer simulations. The core-shell ($CS$) and hollow ($HS$) ellipsoidal microgels were adsorbed at the water/oil interface from the water phase. It was assumed in earlier studies that $CS-165$ microgels have a constant shell thickness, which means that the thickness of the shell at the top and the sides of the core are the same.\cite{Nickel2019,Nickel2021} However, this may not be the case. In order to get a deeper understanding on the correlation between the thickness anisotropy of the shell and spreading of the microgel at the interface, three types of core-shell microgels with the same solid core, shell mass and cross-linking density $CS_{2.8}, CS_{2.5}, CS_{2.2}$ as well as corresponding hollow microgels, $HS_{2.8}, HS_{2.5}, HS_{2.2}$ have been designed in silico (Figure \ref{sim_1}A,B). The aspect ratio of the solid core, $L_{core}/d_{core}$ was set to $4.4\pm0.1$ to match the dimensions of the hematite-silica core in the experiments. The aspect ratio of the shells was varied in such a way to preserve the overall number of monomers and cross-links from sample to sample. After equilibration in a swollen state, the overall aspect ratios of the shell determined from the fitting of $P(q)$, $L/d = 2.8\pm0.1,~2.5\pm0.1$ and $2.2\pm0.1$ were determined for the $CS_{2.8}, CS_{2.5}$ and $CS_{2.2}$, respectively (see Supporting information for further details). The sample $CS_{2.2}$ corresponds to the case of the constant shell thickness surrounding the ellipsoidal core. The ratio between the thicknesses at the tip and the sides, $dL/dd$ = 1.1. The $CS_{2.5}$ and $CS_{2.8}$ samples represent the microgels with slight and strong shell thickness anisotropy,  $dL/dd$ = 1.5 and 1.9, respectively (Figure \ref{sim_1}A). For the investigated systems, removing the core does not significantly change the radii of the hollow microgels similarly to the observations for the microgels with the thin shell\cite{Nickel2019} (Figure \ref{sim_1}B). The higher shell thickness at the tip of the microgel leads to the extra swelling of the shell, Figure S8. Slight increase of the $L/d$ value was observed for the samples $HS_{2.5}$ and $HS_{2.8}$.\\

Equilibrium structures of the $CS$ and $HS$ anisotropic microgels at the decane/water interface are shown on Figure \ref{sim_1}C,D. Decane is a poor solvent and penetration of the oil molecules into the microgels is negligible in comparison with water molecules. The major part of the microgels is located in the water phase, and the narrow polymer/oil interface is slightly bent. After adsorption microgels become significantly stretched due to the minimization of the water/decan surface energy. However, the presence of the solid core and thin, highly cross-linked shell restricts the spreading and preserves anisotropy. Comparing the relative size changes of the $HS$ and $CS$ microgel summarized in Figure \ref{sim_1}E, we found out that the effect of shell anisotropy is also accompanied by increasing the surface area. The highest surface area occupied by the $HS_{2.8}$ in Figure \ref{sim_1}D and gradual decrease of the spreading in a sequence is the following ${HS_{2.8} \rightarrow HS_{2.5} \rightarrow HS_{2.2} \rightarrow CS_{2.8} \rightarrow CS_{2.5} \rightarrow CS_{2.2}}$. Figure \ref{sim_1}G,F presents the simulated aspect ratio of the two microgel types in bulk and at the water/oil interface. When both $CS$ and $HS$ microgels have almost similar aspect ratio in bulk, the influence of the microgel design is more striking at the interface. Interestingly, increasing of the aspect ratio of the microgels lead to the extra stretching of the $CS$ microgels at the tip of the solid core accompanied by the increasing of the shape anisotropy, $L/d$. Thus, in the case of $CS_{2.8}$ microgel, $L/d$ = 3.0$\pm$0.1 which is higher compare to the bulk. On the contrary, etching away of the core lead to the decreasing of $L/d$. In the case of $HS_{2.8}$ microgel, $L/d$ = 2.6$\pm$0.1. This trend is reverse when the shell thickness is comparable at the sides and tips, $HS_{2.2}$ microgels become slighlty more anisotropic than $CS_{2.2}$ microgels.

\begin{figure}[H]
\centering
\includegraphics[width=1\linewidth, trim={0cm 0cm 0cm 0cm},clip]{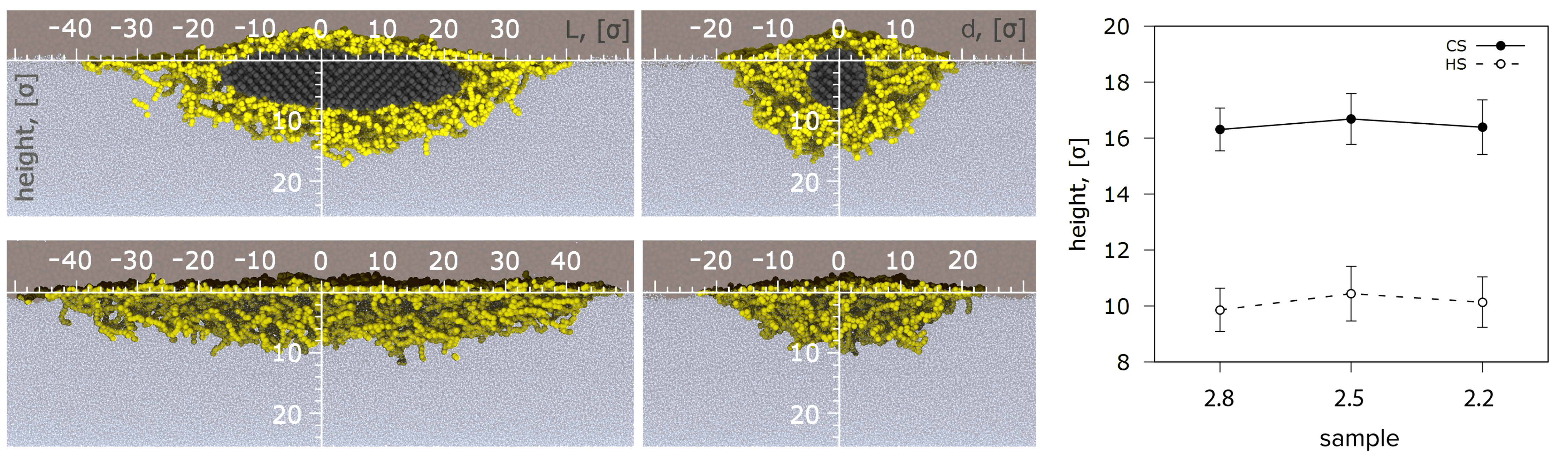}
	\caption{Longitudinal and lateral cross-sections of $CS_{2.2}$ and $HS_{2.2}$ anisotropic microgels at the decane/water interface. Water and decane phases are colored in blue and brown respectively. Maximum heights of the $CS$ (full symbols, solid line) and $HS$ anisotropic microgels (hollow symbols, dashed line) for the different designs are shown on the right.} 
	\label{sim_2}
\end{figure}

The heights of the CS and HS anisotropic microgels for the different design are shown in Figure \ref{sim_2}C. The microgels show a decrease within the height after etching away of the core. In Figure \ref{sim_2}A,B the lateral cross-sections of $CS_{2.2}$ and $HS_{2.2}$ are presented. The cavity is distinguishable, but highly stretched parallel to the interface. The results are in a good agreement with the experimental observations keeping in mind that the experimental height profiles were captured in the dried state. Indeed, due to the solid silica coated hematite core, the relative difference of height between the two microgel types is much more significant in the dried state compared to the swollen one.  \\

Furthermore in the experiments, the length and diameter of $HS-165$ decreased with increasing surface pressure to 456$\pm$56~$\times$~213$\pm$12 which differs from $CS-165$ having similar sizes in the uncompressed region I (355$\pm$40~$\times$~195$\pm$14) and the compressed region V (367$\pm$37~$\times$~186$\pm$14). This observation thus supports the fact that the $HS$ microgels are more deformable at the interface.  Looking back to our simulations, $CS_{2.2}$ and $HS_{2.2}$, corresponding to microgels with approximately the same shell thickness at the sides and tips provide the closest description of the experiments as  summarized in Table \ref{tab_lengthwidth}.

\begin{table}[H]
     \centering
     \small
    \caption{Length, diameter and aspect ratio of the hollow and core-shell microgels obtained from the experimental AFM images and simulations.}
     \begin{tabular}{ccccc}
       \hline
microgel & & length & diameter  & aspect ratio\\
       \hline
$CS-165$ & experiment &  355$\pm$40 nm & 195$\pm$14 nm & 1.82$\pm$0.08  \\
$HS-165$ & experiment & 573$\pm$64 nm & 290$\pm$15 nm & 1.98$\pm$0.12 \\ 
$CS_{2.2}$& simulation & 74.9$\pm$1.8$ \sigma$ & 34.8$\pm$1.4 $\sigma$ & 2.15$\pm$0.1 \\
$HS_{2.2}$ & simulation & 92.2$\pm$1.6$ \sigma$ & 40.8$\pm$1.5 $\sigma$ & 2.26$\pm$0.09 \\
\hline
     \end{tabular}
     \label{tab_lengthwidth}
     \end{table}

The simulation results illustrate the intricate interplay between the microgel design and its spreading at the interface. Due to the presence of the core to which the polymeric network is grafted, the spreading of the $CS$ microgels is strongly constrained. As a consequence, an increase of the shell thickness results in a larger interfacial deformation. This can be observed for $CS_{2.8}$ with a much thicker shell at the tip that becomes slightly more anisotropic at the interface. The dimensions of the microgel play however an important role too. Indeed, spherical microgels with similar crosslinking degree flatten more than larger ones which can be explained by higher relative contribution of the interfacial interactions in comparison with the volume contribution similar to that shown in Eq. (1). Consequently, when the thickness at the side of the anisotropic microgels is comparable to the one at the tip as for $CS_{2.2}$, the microgels deform more laterally resulting in a lower aspect ratio at the interface. 

For $HS$ microgels, the spreading is less constrained, which can be evidenced by the larger occupied interfacial area and decrease height. When looking to polymeric capsules, both the thickness of the shell and dimensions of the cavity are key parameters; a large capsule with thin shell being more deformable. Reducing the relative thickness at the tip has therefore the opposite effect than for the $CS$ microgel, $i.e.$ , the microgels become more anisotropic at least compared to their equivalent $CS$ microgels as shown for $HS_{2.2}$. Further reducing the relative thickness at the tip of our designed $HS$ systems may promote their anisotropic spreading. Finally, we should keep in mind that the dimensions of the cavity may also play an important role as $HS$ microgels can also spread towards the center of the voids and close the cavity. This adds an extra constraint to the interfacial deformation and may explain that the lateral spreading still remains greater than the longitudinal for all the simulated systems.\\
We now discuss the contact line distortion created by a single microgel on a flat decane/water interface. We expect that both microgel design and the presence or absence of the solid core may affect distortion and trigger the capillary forces. We have tracked the position of the microgel perpendicular to the interface and estimated the deformation of the contact line of the liquids for different samples. Figure S12 provides an overview of the position of the microgels relative to the interfaces. The figure shows 3D views of the halves of the CS and HS microgels at the water/decane interface and the results of the surface reconstruction of the interface as a red and white and black gradient color code height maps. The corresponding top view and front view two-dimensional projections can be found in Figure S13. We distinguish reliable differences in the deformation of the contact line between the $CS$ and $HS$ samples. For microgels with a cavity the fluid interface is quite flat. The trench depth is slightly more than 3${\sigma}$ (Figure S14B). The difference between the surface area of the interface with (interface with distortion) and without microgel (undisturbed interface) is shown in Figure S14A. The distortion of the fluid interface is rather small. 
The microgels with solid core create a downward distortion. The polymeric shell tends to spread over the surface, accompanied by downward force that pushes the solid core towards the decane. We observe significant increase both area of deformation and trench depth compared to their counterparts with a cavity, which resulted in the expectation of stronger capillary interactions for the $CS$ anisotropic microgels.

\begin{figure}[ht]
\centering
\includegraphics[width=1\linewidth, trim={0cm 0cm 0cm 0cm},clip]{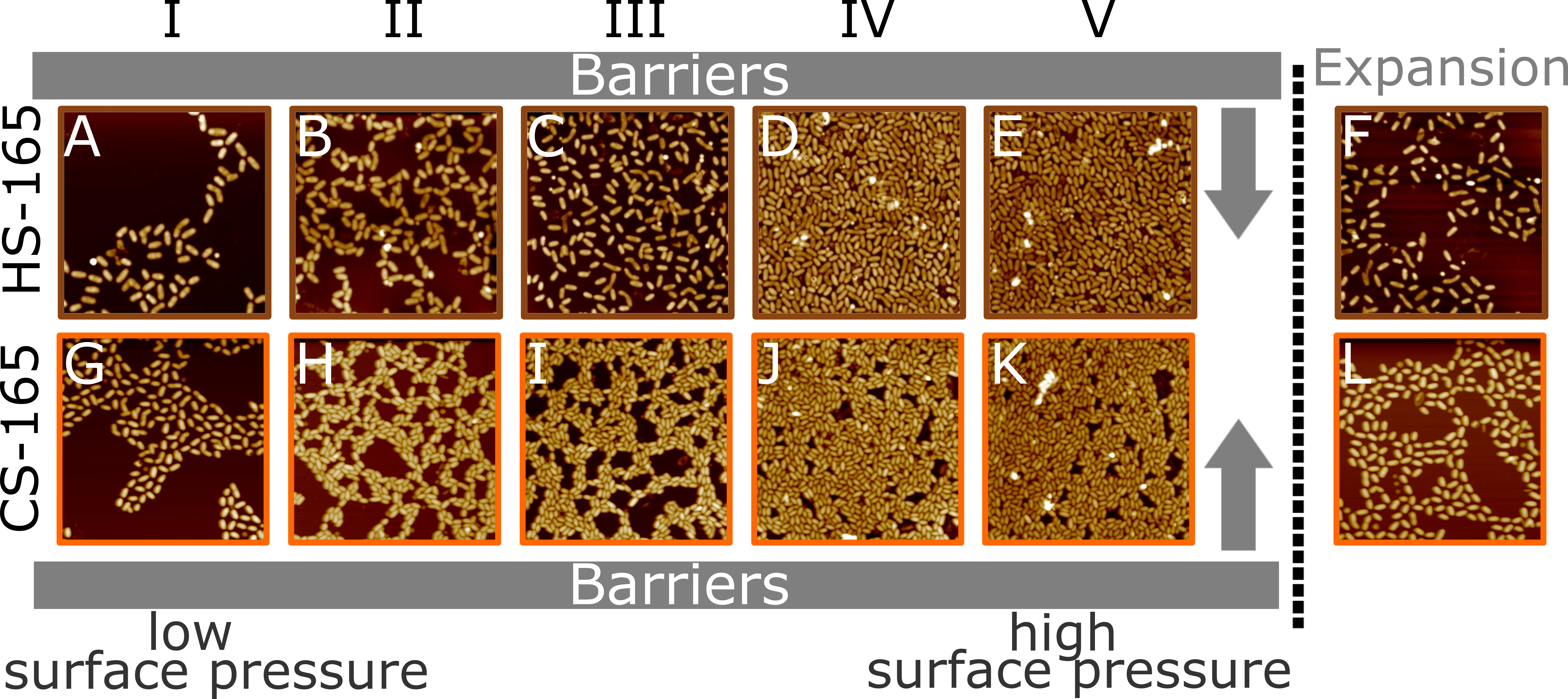}
	\caption{AFM height images of hollow ($HS-165$) (A-F) and core-shell microgels $CS-165$ (G-L) at the five different regions of the compression isotherm and after expansion of the monolayer after the compression. The arrows indicate the direction in which the barriers were compressed in respect to the images. AFM height images are 7.5~µm$\times$7.5~µm. The images of the core-shell microgels are reproduced from Nickel et al. Anisotropic Microgels show their Soft Side. Langmuir 2021, https://doi.org/10.1021/acs.langmuir.1c01748 Copyright 2021 American Chemical Society} 
	\label{HCS_AFM}
\end{figure}

\noindent \textbf{Microstructural analysis of the interfacial assembly.}\\
AFM height micrographs of the deposited monolayers were analyzed to evaluate the different microstructures with increasing compression. Representative micrographs in each of the five regions of the compression isotherm recorded for the hollow and core-shell microgel are displayed in Figure~\ref{HCS_AFM}~A-E and Figure~\ref{HCS_AFM}~H-K, respectively. The direction of the compression with the Langmuir trough barriers is marked with grey arrows. Additionally, AFM images after expansion of the monolayer at the end of the experiment are displayed for the hollow (Fig.~\ref{HCS_AFM}~F) and core-shell (Fig.~\ref{HCS_AFM}~L) microgel. Images F and L exhibit similar heterogeneous structures as observed in region I (Fig.~\ref{HCS_AFM}~A and G)  demonstrating the reversibility of the structures upon reducing the surface pressure. Unlike conventional spherical microgels that usually do distribute homogeneously when the barriers are released, we ascribe this effect to the release of elastic energy stored upon compression overcoming the capillary interaction holding some of the particles at contact.

Both microgels assemble in non-close-packed clusters in region I indicating the presence of attractive capillary interactions balanced by the elasticity of the adsorbed deformed corona independent of the existence of a rigid core for these anisotropic microgels. Note that no clear preferential assembly could be observed as expected for ellipsoidal particles with such lower aspect ratio.\cite{LUO2019} Still, differences within the ordered structures of $HS-165$ and $CS-165$ are clearly visible. The clusters of the core-shell microgels are larger and more evenly packed as compared to the clusters of the hollow microgels which indicates stronger capillary interactions. Spherical core-shell microgels that have a similar diameter as the length of the anisotropic core-shell microgels do not form clusters at low surface pressures.\cite{Rauh2017} This indicates that it is the anisotropic shape of the microgel shell that leads attractive capillary interactions, which might be further strengthened by the extra weight and higher rigidity of the core-shell microgels. 

A similar trend is visible for the other four regions of the compression isotherms. For $CS-165$, the presence of uncovered areas and thus of a highly inhomogenous assembly of microgels at the interface even at the highest surface pressure (Fig.~\ref{HCS_AFM}~K). In contrast, the  nearly homogeneous distribution (without any uncovered areas) of the hollow microgels at high surface pressures (Fig.~\ref{HCS_AFM}~D and E)  indicates weaker capillary interactions and larger compressibility. This could be an additional effect of the possibility of the soft hollow microgels to bend and therefore slightly change their shape to fit into the microgel structures.

\begin{figure}[ht]
\centering
\includegraphics[width=1\linewidth, trim={0cm 0cm 0cm 0cm},clip]{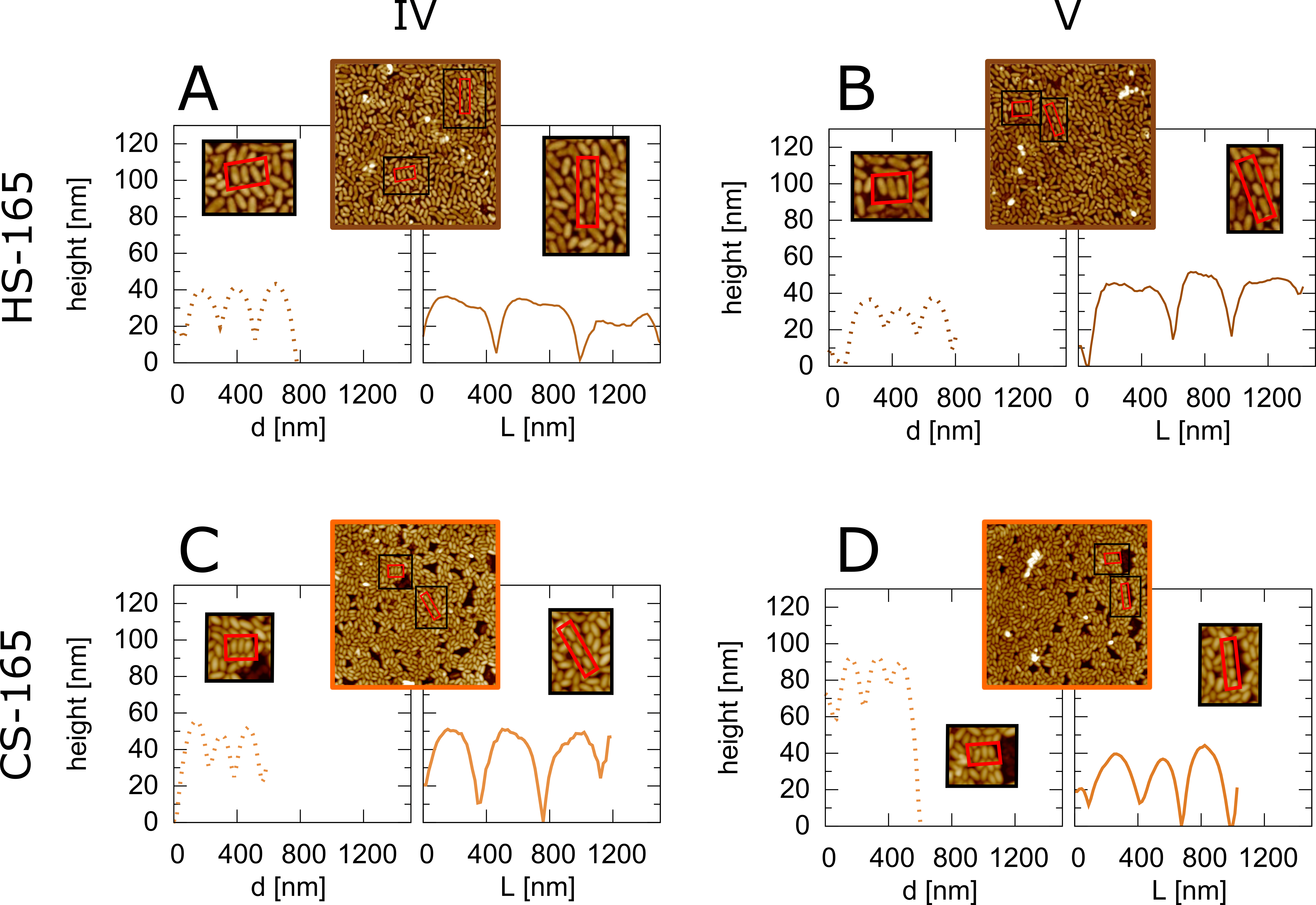}
	\caption{Height profiles determined along the short axis (dotted lines) and long axis (solid line) of three microgels in a $s-s$ and $t-t$ configuration, respectively. The larger micrographs were acquired within region IV and V of the compression isotherm  for $HS-165$ (IV: A;V: B) and $CS-165$ (IV: C; V: D) and have a size of 7.5~µm~$\times$~7.5~µm. The smaller micrographs are cut outs which were enlarged from the larger micrographs as highlighted in them.} 
	\label{HCS_Profil}
\end{figure}

A closer examination of the monolayer topology is shown in Figure~\ref{HCS_Profil} and illustrates height profiles of $s-s$ and $t-t$ clusters for $HS-165$ (Fig.~\ref{HCS_Profil}A, B) and $CS-165$ (Fig.~\ref{HCS_Profil}C, D) within the high compressed regions IV (Fig.~\ref{HCS_Profil}A, C) and V (Fig.~\ref{HCS_Profil}B, D). Even at high surface pressure, the length of the hollow microgels is larger as compared to the core-shell microgels, similarly to what was found at low surface pressure (see Fig.~\ref{HCS_height}).

The same observation holds for the short axis ($d$) of the microgels which are shown as dotted lines in Fig.~\ref{HCS_Profil}. The height profiles of the short axis show similar distinct peaks for $HS-165$ and $CS-165$ in regions IV and V. The maximum height in Fig.~\ref{HCS_Profil}~D for $d$ is larger compared to the other three images. The reason is, that at such high surface coverage the normalization to zero might not be correct when a bare substrate was not accessible. In the case of Fig.~\ref{HCS_Profil}~D, the chosen $s-s$ cluster of microgels lays next to a darker and properly empty surface area. This led to a more realistic determination of the substrate height and values for the maximum height of circa 90~nm still smaller compared to the maximum value of $\approx$120~nm. The height profile of the length of the hollow microgels display similar, rough and more flattened profiles as observed in Fig.~\ref{HCS_height}~(top). The lengths of each microgel decreases as a result of the compression within the Langmuir-Blodgett trough. For $CS-165$, the height profiles for the long axis reveal distinct peaks which larger dimensions compared to the respective short axis and compared to the initial height profile of the long axis in Fig.~\ref{HCS_height}~(bottom). As only exemplary height profiles are chosen, no concrete conclusion can be drawn about the change in size between the regions IV and V as the alignment of the different $s-s$ or $t-t$ clusters is not perfect and especially differs between the slightly different configurations. Hence, drawing a straight line through three microgels does not reproduce the height profile for the individual microgels perfectly . Nevertheless, we can assert that the dimensions of the individual microgels decreased at high $\Pi$ compared to the individual microgels measured at low surface pressure (Fig.~\ref{HCS_height}).

\begin{figure}[ht]
\centering
\includegraphics[width=0.5\linewidth, trim={0cm 0cm 0cm 0cm},clip]{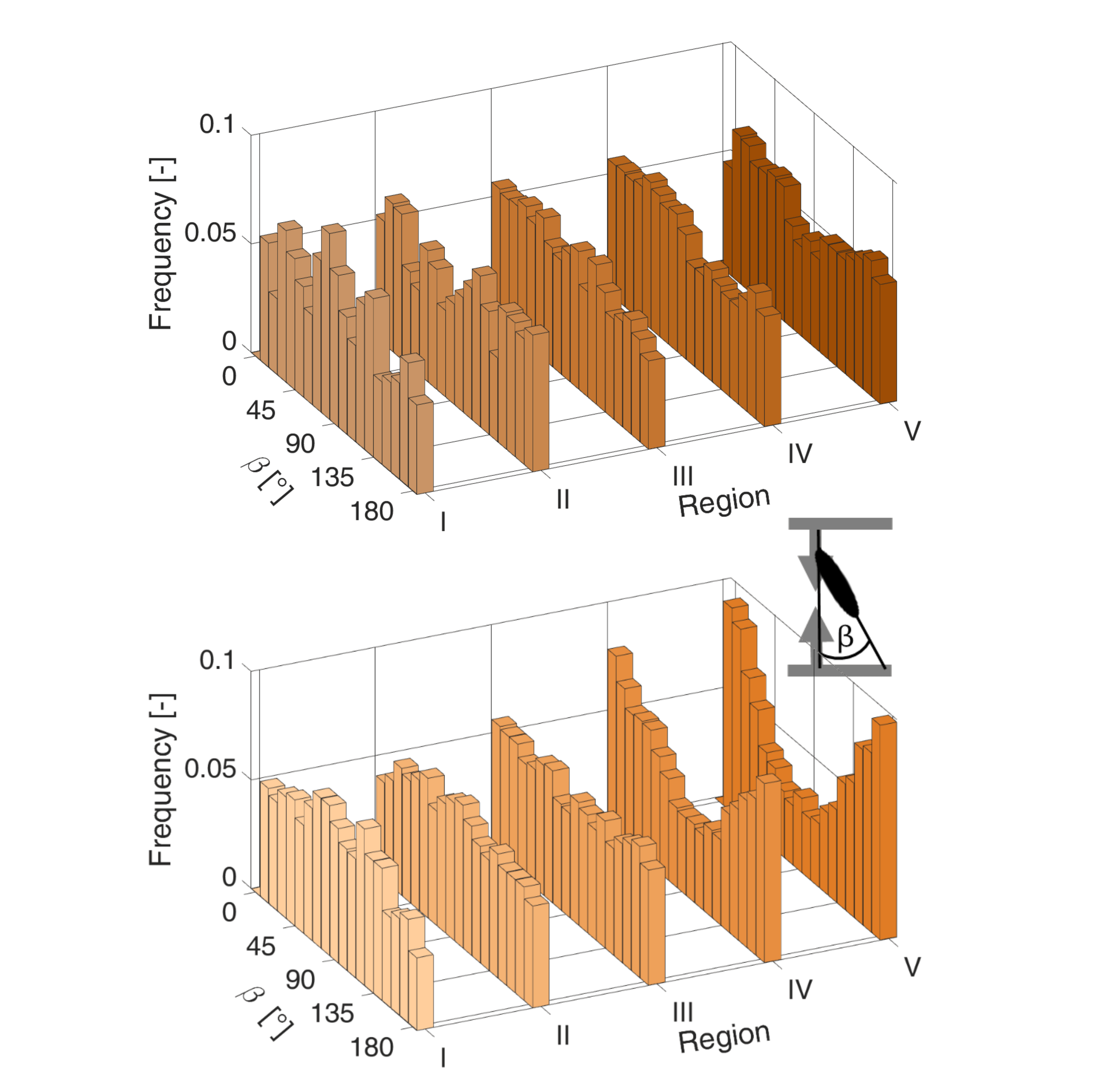}
	\caption{ Development of the orientation of the microgels ($\beta$) of $HS-165$ (top) and $CS-165$ (bottom) for the five different regions within the compression isotherm. Additionally, a definition of the orientation $\beta$ of the microgels is illustrated in the right of the image. The data of the core-shell microgels are reproduced from Nickel et al. Anisotropic Microgels show their Soft Side. Langmuir 2021, https://doi.org/10.1021/acs.langmuir.1c01748 Copyright 2021 American Chemical Society} 
	\label{HCS_Histogramme}
\end{figure}

The orientation of all microgels ($\beta$) compared to the direction of compression with the Langmuir trough barriers, the nearest neighbor distances ($NND$) and the amount of $t-t$ and $s-s$ ordering were computed for both microgels. For this $\beta$ is defined as the angle between the direction of compression with the barriers and the long semi-axis of the microgel cores. A sketch highlighting the definition of $\beta$ in dependency of the direction of compression is shown in Figure~\ref{HCS_Histogramme}.  The obtained values of $\beta$ for the five different regions are reported as histograms in Figure~\ref{HCS_Histogramme}~(top) for $HS-165$ and in Fig.~\ref{HCS_Histogramme}~(bottom) for $CS-165$. No clear preferential direction for any of the five regions could be identified for $HS-165$. The angles 0-60° showed slightly higher frequency values compared to the ones above 90° for region III - V. Conversely, $CS-165$ showed a clear preferential orientation parallel to the direction of compression at regions IV and V.\cite{Nickel2021} When compared to the clear trend observed in Fig.~\ref{HCS_Histogramme}~(bottom) for $CS-165$, $HS-165$ can therefore be accounted as randomly orientated.

Hence, the presence of the hard, in-compressible core has a major influence on the overall orientation of the core-shell microgels at higher compressions as discussed in more detail in our former study.\cite{Nickel2021} The larger stiffness of $CS-165$ microgels was here responsible for the preferential direction observed for the anisotropic core-shell microgels in region IV and V of the compression isotherm. As both systems present a similar aspect ratio at the interface, we conclude that the stiffness of the anisotropic microgels plays a key role in their orientation under the directional compression set by the trough barriers.

In addition, the nearest neighbor distance $NDD$ was computed within the five regions as summarized in Fig.~S3. This analysis first confirms that the average $NDD$ decreased upon compression for the two systems. This effect is attributed to the deformability of the microgels, but also to the promoted $s-s$ assembly at larger compression. Indeed the $NDD$ distributions presented in Fig.~S3 show a broader distribution at low compression. This is particularly the case for $HS-165$ for which the observed clusters do not present a clear preferential configuration at low compression. Increasing the surface pressure then results in a more compact preferential $s-s$ assembly characterized by the narrowing of the distribution and its decrease to shorter distances. The less deformable $CS-165$ present stronger capillary interactions characterized by a preferential $s-s$ assembly already at lower surface compression.\\

\begin{figure}[ht]
\centering
\includegraphics[width=0.4\linewidth, trim={0cm 0cm 0cm 0cm},clip]{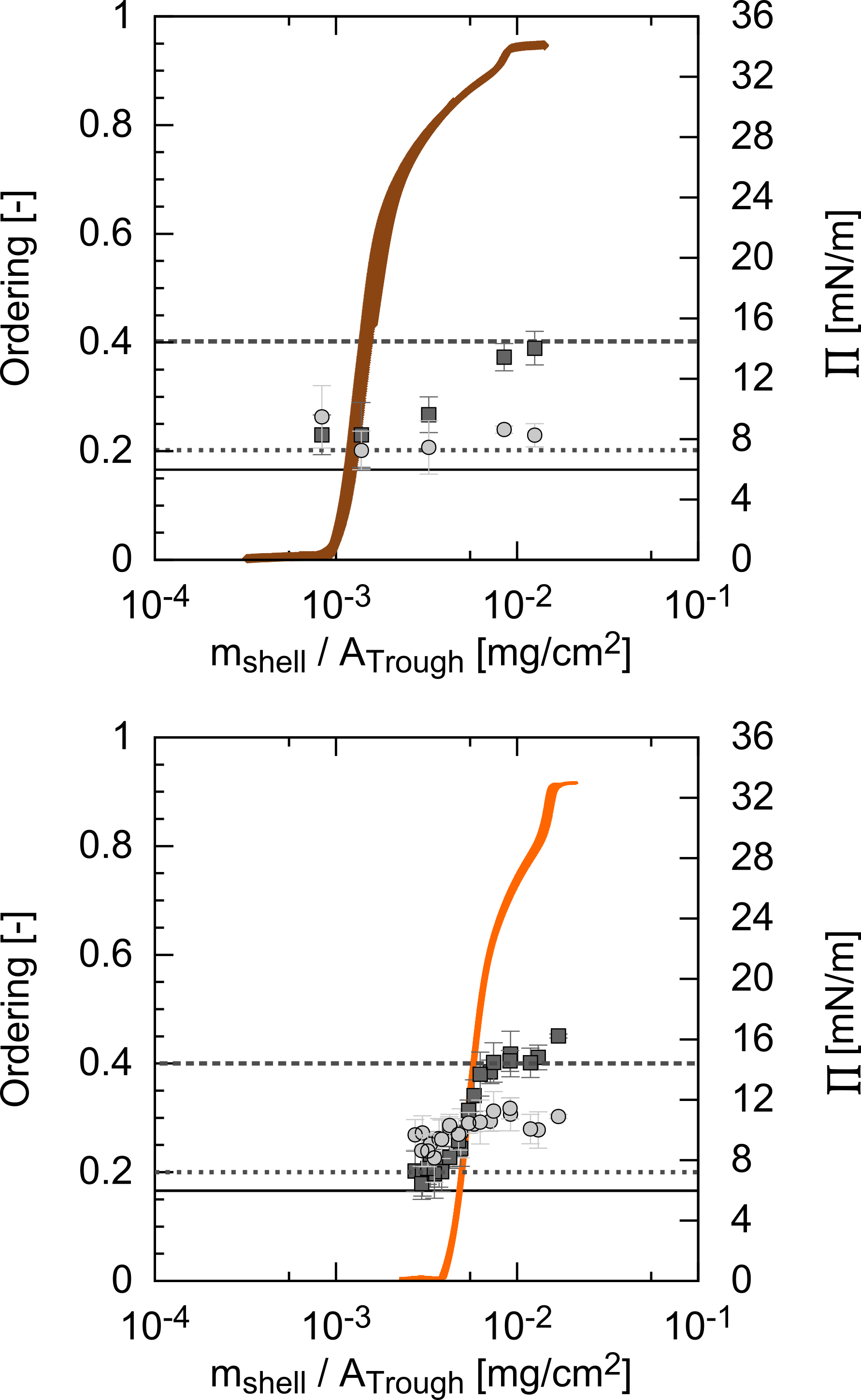}
	\caption{Compression isotherms of $HS-165$ (top) and $CS-165$ (bottom) as function of $m_{shell}/A_{Trough}$ reported together with the amount of $t-t$ (circles) and $s-s$ ordering (squares). The continuous line at 0.167 refers to the value for a random orientation. The dotted lines at 0.2 and 0.4 are a guide to the eye for an easier comparison between the two systems. The data of the core-shell microgels are reproduced from Nickel et al. Anisotropic Microgels show their Soft Side. Langmuir 2021, https://doi.org/10.1021/acs.langmuir.1c01748 Copyright 2021 American Chemical Society} 
	\label{HCS_Ordering}
\end{figure}

To obtain more detailed information about the short range ordering of the microgels, the $s-s$ and $t-t$ ordering is determined similar to ref. \cite{Nickel2021}. We refer to this study for more details on this analysis. In a nutshell, we considered neighboring microgels presenting the same orientation $\pm$ 15°. An $s-s$ configuration was identified if the nearest neighbor center was located at the side of a microgel with a tolerance of $\pm$ 30°. Similarly, a $t-t$ configuration was considered when the center of the nearest neighbor lay in the direction of the long axis $\pm$ 25°. A random orientation is characterized here by a value in the order of 0.167. Hence, only the orientations of the nearest neighbors were compared and related to the different configurations. \\
Figure~\ref{HCS_Ordering} presents the compression isotherms of hollow (Fig.~\ref{HCS_Ordering}~top) and core-shell (Fig.~\ref{HCS_Ordering}~bottom) microgels together with their respective $t-t$ (circles) and $s-s$ (squares) ordering with increasing surface compression. For $HS-165$ (Fig.~\ref{HCS_Ordering}~top), the $s-s$ ordering increases for higher $m_{shell}/A_{area}$ while the $t-t$ ordering remains constant. This increase in $s-s$ ordering appears from region III to region IV in the compression isotherm. 

In contrast to $HS-165$, both the $t-t$ and $s-s$ ordering increase at higher surface pressure for $CS-165$. Nevertheless, the increase in $s-s$ ordering is larger compared to the increase in $t-t$ ordering leading to a higher $s-s$ ordering of $CS-165$ in region III, IV and V.\\
Hence, the preferential $s-s$ ordering is a result of the elliptical shape of the microgel and only slightly influenced by the hard elliptical core. Even in the compressed region V, the elliptical shape of the hollow microgels is preserved leading to $s-s$ ordering indicating shape-dependent short range capillary interactions as also observed for hard ellipsoids.\cite{Botto2012} The elliptical core slightly enhances this trend as the $s-s$ ordering of $HS-165$ goes up to 0.4 and the $s-s$ ordering of $CS-165$ increases to slightly higher values. Additionally, the preferred $s-s$ ordering appears already in region I\hspace{0.05cm}I for $CS-165$ which results in $\sim$0.4 $s-s$ ordering already in region III. This is different for $HS-165$ where the high values of $\sim$0.4 $s-s$ ordering are observed in region IV and V only which is the region where no prominent clustering was observed. To conclude, the hard core has an influence on the nearest neighbor ordering but the general type of order is determined by the elliptical shape of the microgels in both cases.

\section{Conclusions}

In this work, we investigated the influence of a hard elliptical core on the behavior of anisotropic microgels at a decane-water interface. For this, we prepared first a core-shell microgel with a thin and stiff polymer shell synthesized around an anisotropically shaped inorganic core and then prepared a hollow, anisotropic microgel via etching the core. Comparing the interfacial behavior of core-shell $CS$ and hollow shell $HS$ anisotropic microgels thus having the same shell properties, the examination of the compression isotherms confirms the promoted deformability of the $HS$ system. Both systems maintain an anisotropic shape at the interface with an aspect ratio in the order of 2. 

Supported by computer simulations, we highlight how the combination of shape anisotropy and microgel architecture influences the microgel deformation at the interface. We first confirmed that hollow microgels spread more at the interface than the stiffer core-shell microgels. A closer investigation of the interfacial deformation revealed that the local shell thickness plays a crucial role and $CS$ microgels with a thicker shell at the tip were found to become more anisotropic at interface than in bulk. For $HS$ microgels, similar to capsules, both the local thickness and dimensions of the cavity are key parameters. For our design it turns out that hollow microgels are then conversely to core-shell microgels becoming more elongated when the lateral shell thickness is in the order of the longitudinal one. From the trend, we can assume that this effect will be amplified when the lateral thickness exceeds the longitudinal one. We believe that these conclusions will contribute to a better understanding of the interfacial deformation of soft anisotropic systems with more complex architectures and may be relevant for instance for biological systems such as cells or virus capsids. In addition, our results are expected to motivate the rational design of soft anisotropic objects with a preprogrammed interfacial deformation set by the dimensions of the core/cavity and the softness and local thickness of the polymeric shell.

Interestingly, hollow and core-shell anisotropic microgels were both found to exhibit capillary interactions, which appeared weaker for the more deformable hollow microgels. In line with our former study on anisotropic core-shell microgels,\cite{Nickel2021} it illustrates the importance of the anisotropy and softness on the interfacial assembly and may be employed to create defined nanostructured materials. In this case, more elongated objects with a more preferential side to side self-assembly, such as stretched core-shell microgels\cite{Crassous2014,Honda2019} might be interesting to create more defined assemblies. 

We further demonstrate the influence of monolayer compression on the interfacial ordering. While both systems build up a preferential side to side ordering at larger compression, as revealed by the formation of small nematic domains, stiffer anisotropic core-shell microgels additionally tend to align parallel to the direction of the compression set by the barriers, whereas softer hollow microgels did not. 

Both microgel monolayers are highly compressible allowing to finely tune the nearest neighbor distance between the microgels. Similar to spherical hollow microgels, hollow anisotropic microgels are for more compressible than their core-shell counterparts. We believe our results will allow to extent the use of microgel monolayers for nanolithography to more complex assemblies. Hereby, deposited monolayers of soft microgels were used as mask for the growth of nanowires to create photonic materials. Due to their compressibility, microgels have the advantage to allow the fine control of the nanowire dispacing.\cite{Rey2016_NL,Scheidegger2017} 

The next challenge for anisotropic microgels is to promote their ordering into large nematic, smectic or even crystalline domains at the interface. More complex lattices may in the future be achieved that could polarize and guide the light, which further motivates the control of the domain orientation. Our results, as proofs of concept, assert the potential of anisotropic microgels for such applications but also the challenges it represents: first from the synthetic viewpoint to create highly monodisperse anisotropic microgels with a proper design and second in respect to their interfacial assembly related to their anisotropy and deformability to properly balance capillary interactions and allow long-range ordering at higher compression as well as domain alignment.

\begin{acknowledgement}
We gratefully acknowledge financial support from the Deutsche Forschungsgemeinschaft within SFB 985 “Functional Microgels and Microgel Systems”. We thank Timon Kratzenberg for the help with the experiments on the Langmuir-Blodgett trough.
\end{acknowledgement}

\begin{suppinfo}
The following files are available free of charge.

\begin{itemize}
\item The SI contains detailed information about the image analysis procedure of the AFM images, the compression modulus obtained in the experiments, the nearest neighbor distances at different compression regions in the compression isotherm and the original data of the compression isotherms. Furthermore, the  SI contains the details of the model of the microgels used in simulation as well as the microgels characterization in bulk, including swelling curves, form factor profiles of the microgels P(q), their sizes and shape anisotropy parameters in a swollen and collapsed sates. 

\end{itemize}

\end{suppinfo}

 \bibliography{main} 
 
  \newpage
 \begin{figure}[ht]
\centering
\includegraphics[width=8.25cm, trim={0cm 0cm 0cm 0cm},clip]{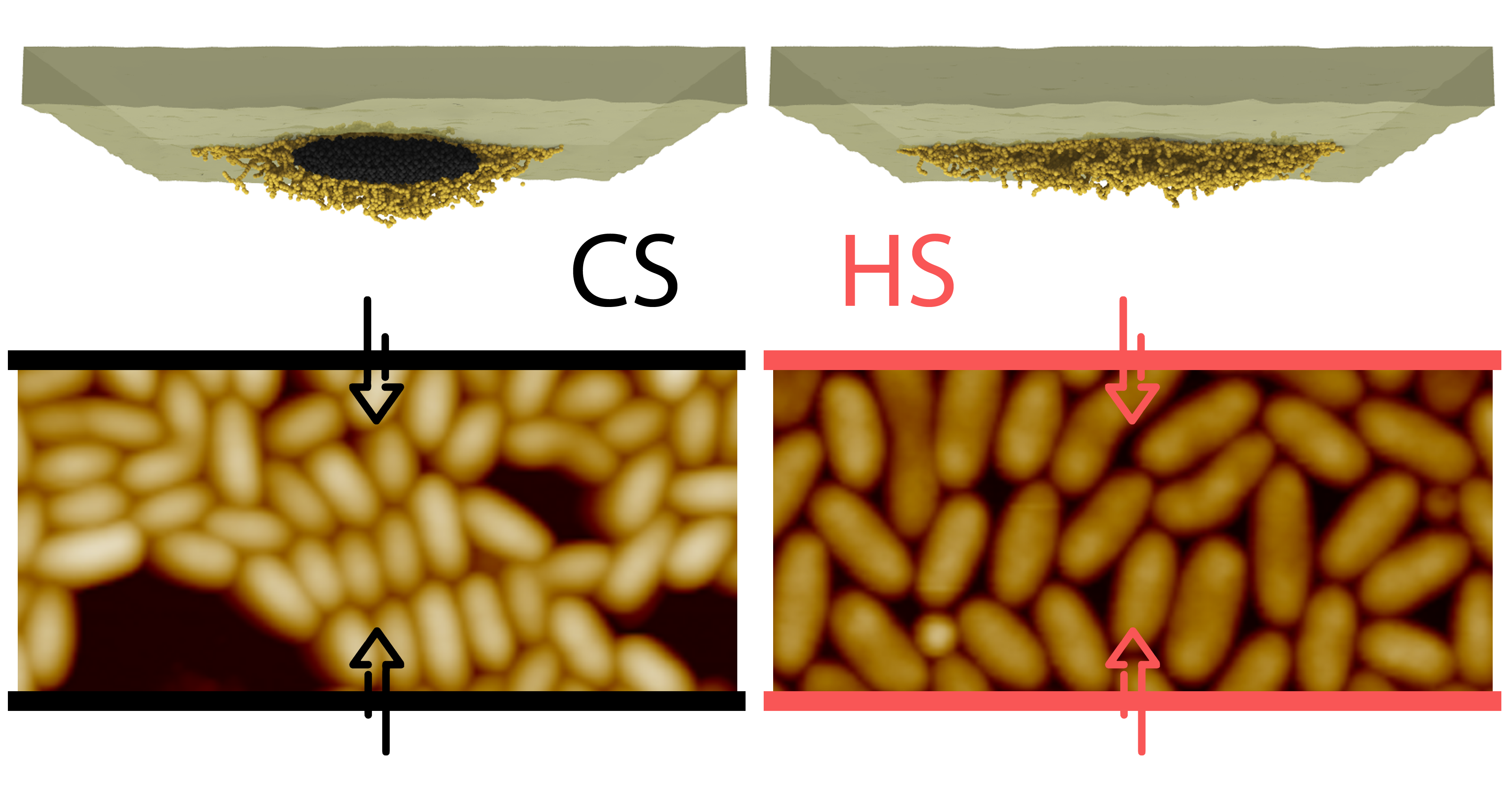}
	\caption{TOC Graphic}
\end{figure}

\end{document}


\noindent \textbf{Image Analysis}\\
The program NanoScope Analysis 1.9 was used for the acquisition and leveling of the AFM micrographs. The processed AFM images were analysed with an updated version of the MATLAB routine of Bochenek et al.\textsuperscript{S1,S2} where the anisotropic microgel position, orientation, dimensions, eccentricity and mean brightness were obtained using the regionprops MATLAB function. Before the microgels could be localized, the images were treated by applying a band-pass filter and a zero crossing edge detection. The AFM height images of $CS-165$ show a sharp decrease in height at the edges of the core which cannot be compressed or deformed as it will be discussed later on. This feature can be tracked with the MATLAB script which can easily distinct between individual microgels by their individual cores for such core-shell microgels. This feature is not existing for the hollow microgels making the detection of the elliptical microgels, which can be observed with bare eye, not possible with the MATLAB script. This problem increases with increasing surface pressure as the distance between the microgels reduces and the microgels become indistinguishable for the script. To overcome this issue, the AFM micrographs of $HS-165$ need to be pre-treated manually to identify the individual ellipsoidal microgels by drawing an elliptical core to the center of the hollow microgels. Such images could be used for the analysis with the previously described MATLAB script. Hence, the investigation of the same amount of images investigated for $CS-165$ was hindered for $HS-165$. Thus, the evaluation for $HS-165$ is conducted for one value for $m_{shell}/A_{Trough}$ for each of the regions within the isotherm. The evaluation for each point is based on six AFM height images for the region I\hspace{0.05cm}I - V and based on four images for region I. \\
The dimensions and ordering results obtained from the analysis of the AFM images are summarized in table~\ref{tab_resultsAFM}.

\begin{table}[H]
     \centering
    \caption{Summary of the results obtained from the AFM images.}
     \begin{tabular}{ccc}
     \hline
     & $CS-165$ & $HS-165$\\
       \hline
     dimensions$_{low~compression}$ & (355$\pm$40~$\times$~195$\pm$14) nm & (573$\pm$64~$\times$~290$\pm$15) nm\\
     dimensions$_{high~compression}$ & (367$\pm$37~$\times$~186$\pm$14) nm& (456$\pm$56)~nm$\times$~(213$\pm$12) nm\\
     t-t$_{low~compression}$ & 0.27 $\pm$ 0.03 & 0.20 $\pm$	0.04 \\
     t-t$_{high~compression}$ & 0.30 $\pm$ 0.02 & 0.23 $\pm$	0.02\\
     s-s$_{low~compression}$ & 0.20	$\pm$ 0.04 & 0.23 $\pm$	0.06\\
     s-s$_{high~compression}$ & 0.45 $\pm$ 0.01 & 0.39 $\pm$	0.03\\
     \end{tabular}
     \label{tab_resultsAFM}
     \end{table}

\begin{figure}[H]
\centering
\includegraphics[width=0.45\linewidth, trim={0cm 0cm 0cm 0cm},clip]{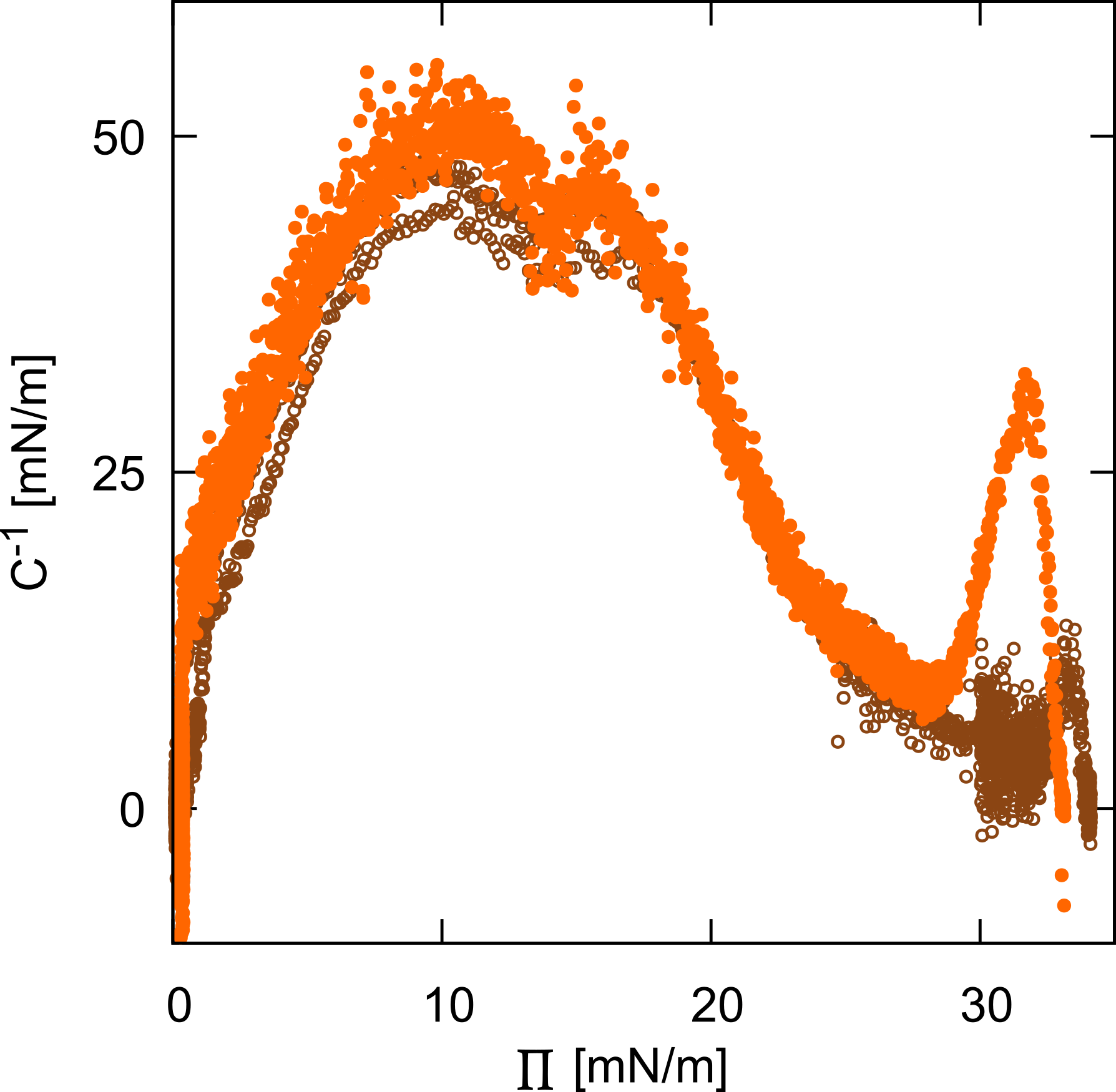}
	\caption{Compression modulus ($C^{-1}$) of $HS-165$ (brown, open circles) and $CS-165$ (orange, closed circles) depending on surface pressure. The displayed values are averages for five points each. Not the full data set of $HS-165$ is shown as a result of the large spread of the data for high surfaces pressures. The averaged full data set is shown in Figure~\ref{HCS_CompressAll} in the appendix.} 
	\label{HCS_Compress}
\end{figure}

\textbf{Compression modulus}\\
Figure~\ref{HCS_Compress} illustrates the averaged compression modulus of $HS-165$ (brown, open circles) and $CS-165$ (orange, closed circles) depending on the corresponding surface pressures.

$CS-165$ has two clear maxima in the compression modulus while for $HS-165$ the second maximum is less pronounced as the data scatters for high surface pressure and the maximum value for $C^{-1}$ is smaller for $HS-165$ compared to $CS-165$. The decrease in the maximum value for the compression modulus for the second peak is a result of the etching of the core.\textsuperscript{S3} Due to the incompressible core, the core-shell microgels show a higher elasticity compared to the hollow microgels for the second peak where the relevance of the core increases.

\begin{figure}[H]
\centering
\includegraphics[width=0.45\linewidth, trim={0cm 0cm 0cm 0cm},clip]{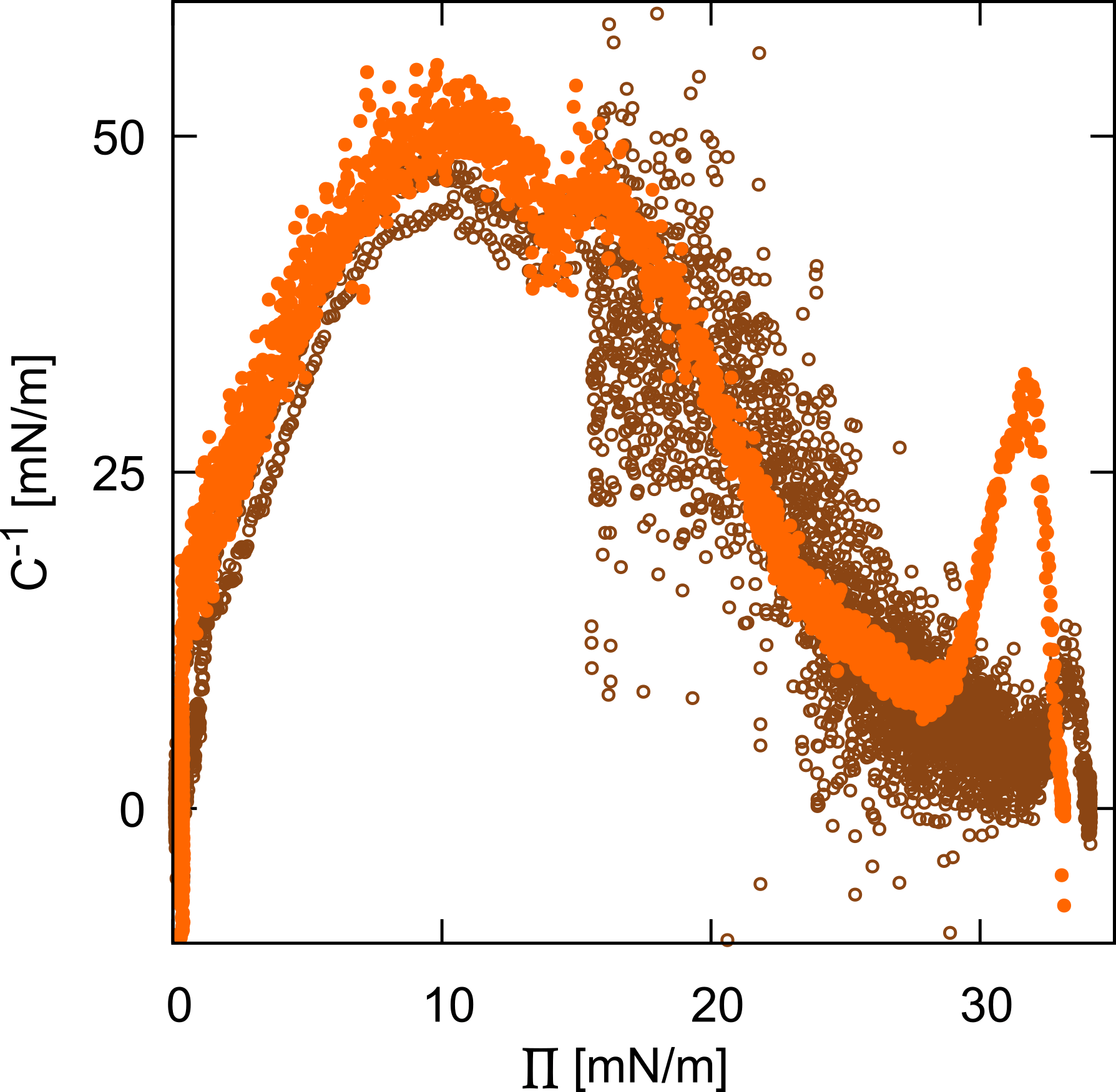}
	\caption{Compression modulus ($C^{-1}$) of $HS-165$ (brown, open circles) and $CS-165$ (orange, closed circles) depending on surface pressure. The displayed values are averages for five points each.} 
	\label{HCS_CompressAll}
\end{figure}

\begin{figure}[H]
\centering
\includegraphics[width=0.6\linewidth, trim={0cm 0cm 0cm 0cm},clip]{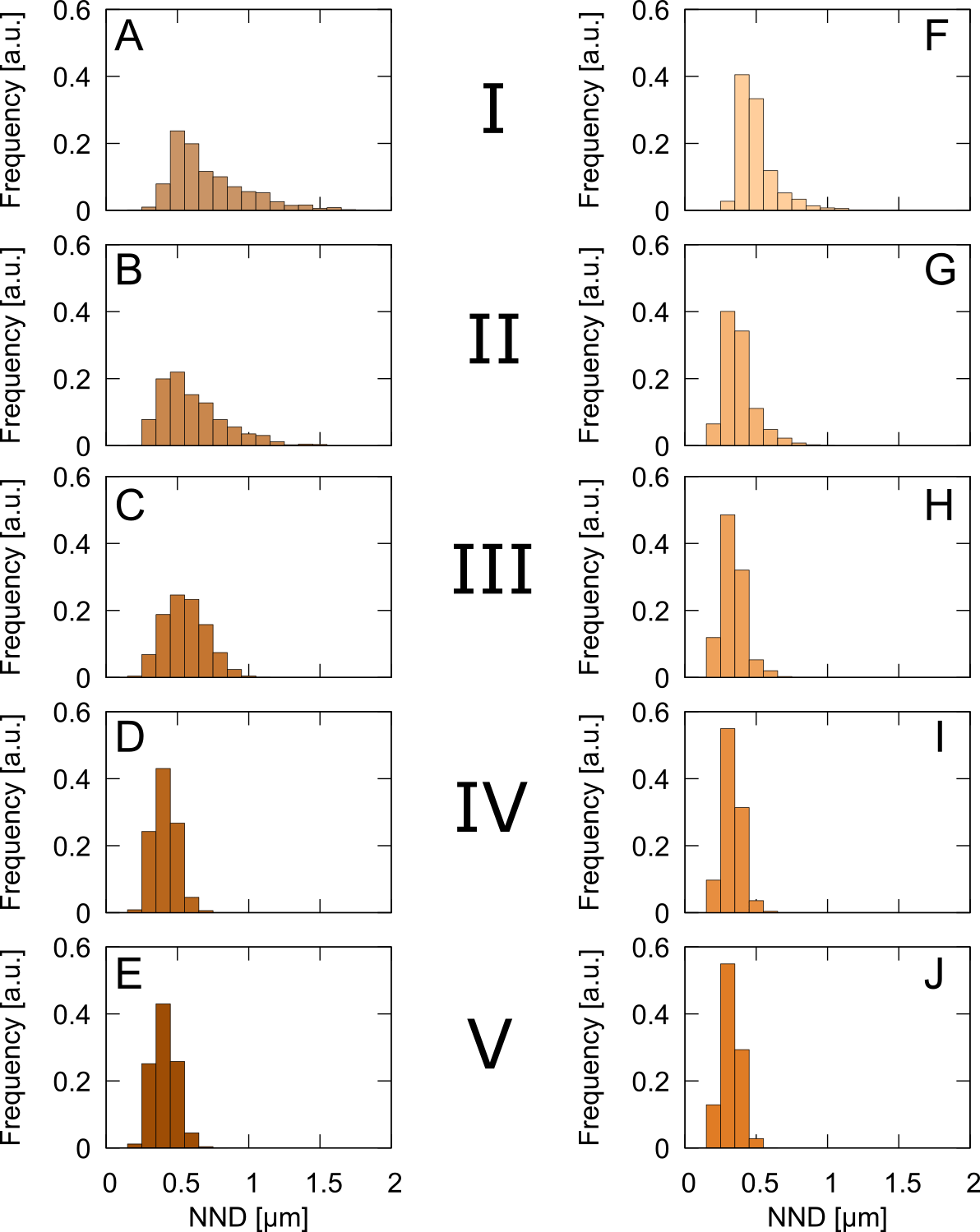}
	\caption{Development of the histograms of the nearest neighbor distances (NND) for $HS-165$ (A-E) and $CS-165$ (F-J) for each of the five regions in the compression isotherm. The data of the core-shell microgels are reproduced from Nickel et al. Anisotropic Microgels show their Soft Side. Langmuir 2021, https://doi.org/10.1021/acs.langmuir.1c01748 Copyright 2021 American Chemical Society} 
	\label{HCS_NND}
\end{figure}

\textbf{Nearest neighbor distances}\\
The development of the NND are illustrated as histograms in Figure~\ref{HCS_NND}~A-E for $HS-165$ and in Figure~\ref{HCS_NND}~F-J for $CS-165$. For each of the regions from the compression isotherm a separate image shows the distribution of the nearest neighbor distance from low surface pressure (A and F) to high surface pressure (E and J). The colors are identical to Figure~7 as the same position within the compression isotherm are used for the histograms of the orientation in Figure~7 and the nearest neighbor distances.\\
$HS-165$ and $CS-165$ show a decrease in NND when coming from region I to region IV. This is expected for microgels when the available surface area is decreased leading to an increase in surface pressure within the Langmuir-Blodgett trough. Furthermore, the change in NND is larger for $HS-165$ when coming from region I to region V compared to the small change in NND observed for $CS-165$ between region I and region V.\\

\begin{figure}[H]
\centering
\includegraphics[width=0.6\linewidth, trim={0cm 0cm 0cm 0cm},clip]{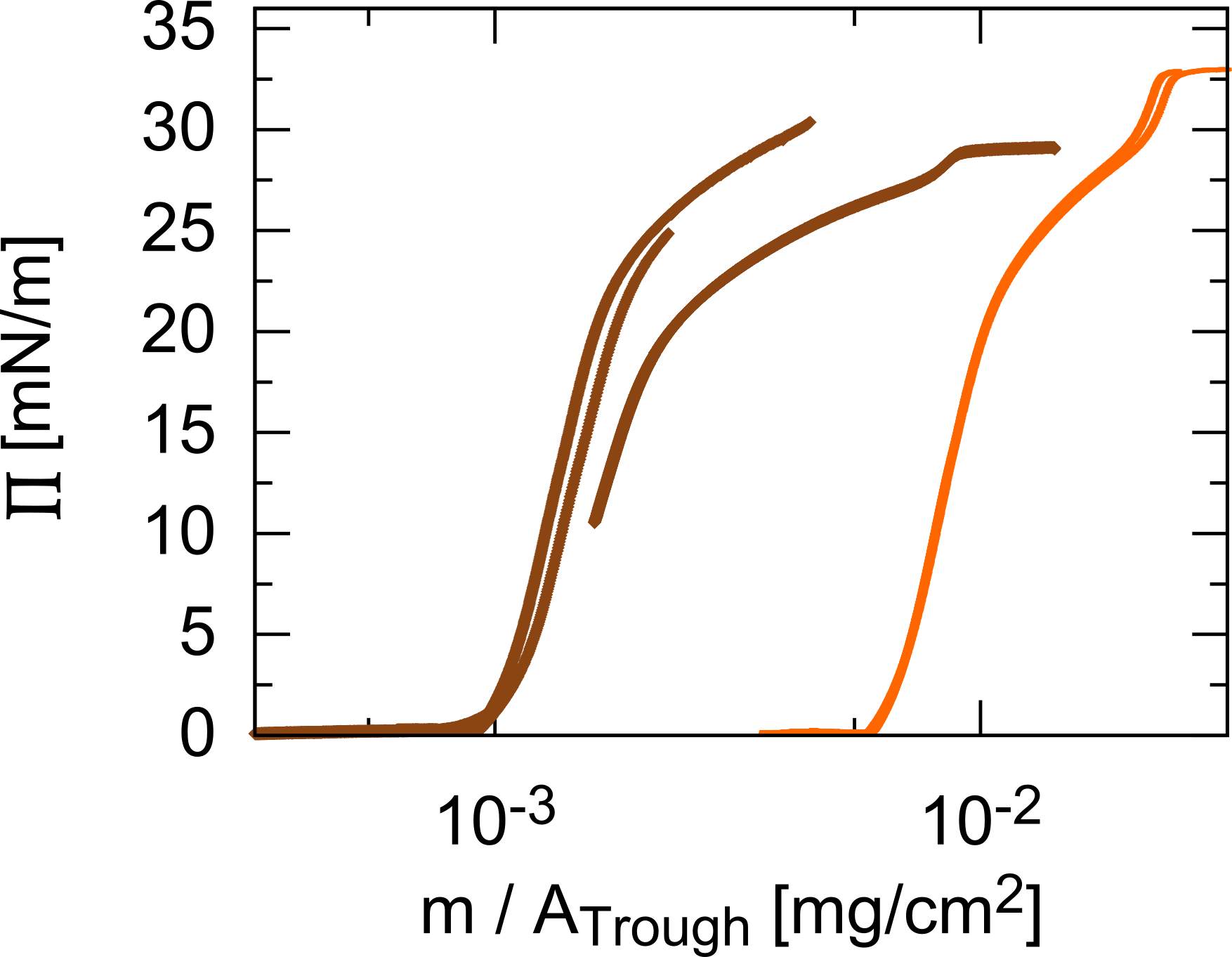}
	\caption{Original data of the compression isotherms of $HS-165$ (brown) and $CS-165$ (orange) in dependence of the mass of the shell divided by the trough area. The compression isotherms have five regions each but differ in the shape and position as a result of the etching away of the core. The data of the core-shell microgels are reproduced from Nickel et al. Anisotropic Microgels show their Soft Side. Langmuir 2021, https://doi.org/10.1021/acs.langmuir.1c01748 Copyright 2021 American Chemical Society} 
	\label{HCS_Isothermen_original}
\end{figure}

Nevertheless, the magnitude of the change in NND is different. $HS-165$ has a broader histogram with higher NND compared to $CS-165$ in region I. In all regions, the NND histograms of $CS-165$ are narrow compared to the affiliated NND histograms of $HS-165$. Those differences in the histograms of the NND are explained by two effects. First, $HS-165$ spreads more at the interface compared to $CS-165$. This leads to a larger occupied area at the interface and, as a result, to larger values for the NND. Second, the capillary interactions for $CS-165$ are promoted as a result of the weight of the core. This leads to a preferential value for the NND even in the dilute state. Hence, the histogram is thinner and shows smaller NNDs. Additionally, the change in the most probable value for the NND is less for $CS-165$, as the ellipsoidal microgels have small distances already in the diluted region because of the attractive shape-dependent capillary interactions which leads to clustering.\\
\textbf{Original data of the compression isotherm}\\
Figure~\ref{HCS_Isothermen_original} shows the original compression isotherm of $HS-165$ and $CS-165$ obtained from different Langmuir-Blodgett trough experiments. While the compression isotherms of $CS-165$ complement each other, the compression isotherms of $HS-165$ do not lie on top of each other.

\textbf{Computer simulations details}

\noindent \textit{Size-limit effect.}
If we address to the collapse of a microgel in solution, the free energy of a single microgel can be written as a sum of two terms:
\begin{equation}
  \textcolor{black}{  F = f V + \gamma S = V [f + \gamma S / V]}
    	\label{simeqF}
\end{equation}
The first one is the volume contribution (proportional to the volume $V$ of the microgel); $f$ being the free energy density and includes subchain elasticity, short-range interactions of polymer with the solvent, solvent entropy, etc. The second term is a surface free energy of the microgel in a poor solvent; $\gamma$ being the surface tension coefficient. This term is proportional to the microgel surface area, $S$. The density $f$ does not depend on the microgel size, however $S/V$ is inversely proportional to the characteristic size of the microgel. In case of the spherical microgel $S/V=3/R$, where $R$ is the microgel radius. Such dependence means that the smaller the microgel, the higher the contribution of the surface tension to the total free energy. And vice versa: in the thermodynamic limit $R\to\infty$ (the limit of the macroscopic gel), the surface does not play a role on the microgel collapse. In computer simulations, because of the limitation in the size of the simulation box (number of the beads), the characteristic sizes of the microgels are much smaller than in the experiments (tens of nanometers). Therefore, the modeling of the microgels, which have a size of hundreds of nanometers, requires a coarse-graining procedure leading to the effective reduction of gel size resulting in a recalculation of all the mechanical properties of the microgels. In the case of small microgels, the system becomes very sensitive to the disbalance between surface/volume terms, which is especially noticeable for oblong or hollow objects at the same coarse-grained level.

\noindent \textit{Method.} The DPD uses a simplified description of the particles (coarse-grained approach).\textsuperscript{S4} In the DPD formalism, the bead is not just representing a single atom or a molecule of a real fluid, but rather a group of molecules localized in a certain volume and moving coherently. The beads simply follow Newton’s equations of particle motion:
\begin{equation}
  m_{i} \ddot{\boldsymbol{r}}_{i} = m_{i} \dot{\boldsymbol{v}}_{i} =  \sum\nolimits_{i\neq j} (\boldsymbol{F}_{ij}^{C} + \boldsymbol{F}_{ij}^{D} + \boldsymbol{F}_{ij}^{R}),
    	\label{simeq1}
\end{equation}
\noindent where $m_{i}$, $\boldsymbol{r_{i}}$, and $\boldsymbol{v_{i}}$ represent the mass, position vector, and velocity vector of a particle i, ${i \in (1..N)}$, respectively, $N$ is the total number of particles, the superposed dot denotes a time derivative, the forces on the right side represent the conservative (subscript C), the dissipative (subscript D), and the random (subscript R) forces, respectively. \\
One of the most important parameters that can be tuned within a DPD simulation are the interaction parameters $a_{ii}$ and $a_{ij}$ describing repulsions between similar and dissimilar unbonded DPD beads, respectively. The parameters $a_{ij}$ are expressed in units of $k_{B}T/r_{c}$, where $r_{c}$ is the cutoff distance, which defines the range of interaction between two beads. There are several methodologies to choose it to describe phase separating fluids or fluid mixtures.\textsuperscript{S5-S8} Following the parametrization method employed by Groot and Warren\textsuperscript{S9} and Maiti \& McGrother\textsuperscript{S5}, one could possible to represent a DPD particle as a single or multiple, $N_m$ water molecules. Based on the assumption of $a_{ii}=a_{jj}$, $r_{c_{i}} = r_{c_{j}}$ and maintaining the constant density in the box, the DPD $a_{ij}$ parameters could be related to the Flory-Huggins theory:
\begin{equation}
  a_{ij} = ( a_{ii}+3.27\chi_{ij})k_BT/r_c,
    	\label{simeq2}
\end{equation}
\noindent To preserve the compressibility, $\kappa^{-1}$, of water at different coarse-grained levels, $a_{ij}$ has to be related to the density, $\rho_{DPD}$,  within the simulation box and to the degree of coarse-graining, $N_m$, as
\begin{equation}
  a_{ii} = k_BT \frac{\kappa^{-1}N_m - 1}{2\alpha\rho_{DPD}},
    	\label{simeq3}
\end{equation}
where $\alpha$ is considered to be a constant equal to 0.1 and independent of $\rho_{DPD}$ and $a_{ii}$. We set the total number density of the system as $\rho_{DPD}=3r_c^{-3}$. There are four different types of particles in our system: water (W), decane (D), the beads forming polymer NIPAM microgel shell (S) and solid core (C). Thus we have to set $a_{ij}$ where $i$ and ${j \in (W, D, S, C)}$ to establish the interactions between the different constituents.\\

\noindent \textit{Decan/water interface.}To reproduce the decane/water mixture we set $a_{ij}$ in such a way to follow the experimental interfacial tension coefficient  52.33 mN/m.\textsuperscript{S10} We treated the decane as a single bead in the water/oil system similarly to Ref.\textsuperscript{S11}. In such representation the $a_{WW}=a_{DD} = 25$ $k_BT/r_c$. The interaction parameter between the liquids can be calculated based on the relation between the Flory-Huggins $\chi$ parameters and the activity coefficient of water. In the limit of infinite dilution: 
\begin{equation}
  \chi_{ij}^\infty = \ln{\gamma_{i}^\infty}+ \ln{v_{ij}}-(1-\frac{1}{v_{ij}}),
    	\label{simeq4}
\end{equation}
where $\gamma_{i}^\infty$ is the infinite dilution activity coefficient of water in decane and $v_{ij}$ is the ratio of the molecular volume of the decane $v_j$ to water $v_i$, ($i=W$, $j=D$). The infinite dilution activity coefficient of water in decane as a function of temperature is presented in Table~\ref{tab_dilution_act}.
\begin{table}[H]
     \centering
     \small
    \caption{Infinite dilution activity coefficients ${\ln{\gamma_{W}^\infty}}$ of water in decan\textsuperscript{S11} and interaction parameters $a_{WD}$ as a function of temperature over the range from $20^\circ C$ to $55^\circ C$. The volume of water, $v_W$ = 25.58 {\AA}$^3$, and $R^2$ is the fitting correlation coefficient. }
     \begin{tabular}{ccccc}
       \hline
${\ln{\gamma_{W}^\infty}}$ & $v_D$ ({\AA}$^3$) & ${v_{WD}}$ & ${a_{WD}}$  & $R^2$\\
       \hline
-0.0095T$^\circ$C+6.808 & 233.02 & 9.11 & -0.0331 T$^\circ$C + 53.447 & 0.9999 \\ 
 \hline
     \end{tabular}
     \label{tab_dilution_act}
     \end{table}
The interfacial tensions obtained from DPD at $20^\circ C$ is 51.9 mN/m which is in good agreement with the experimental value 52.33 mN/m.\textsuperscript{S10}\\

\begin{figure}[ht]
\centering
\includegraphics[width=0.9\linewidth, trim={0cm 0cm 0cm 0cm},clip]{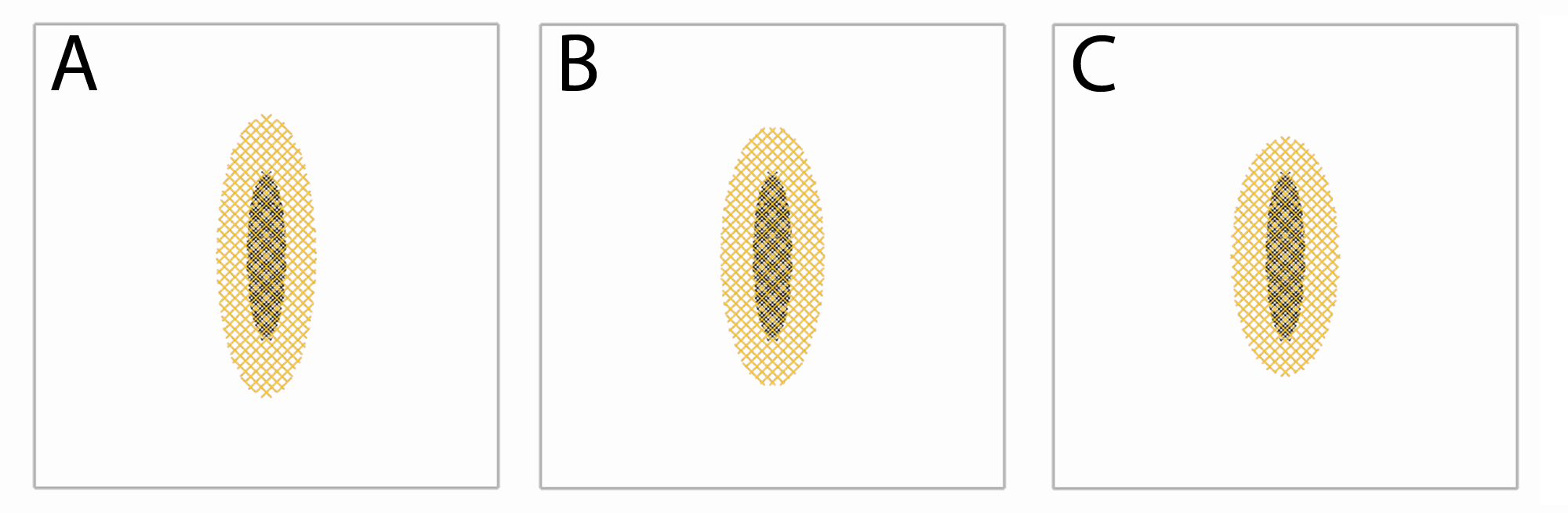}
	\caption{The images of the models of the core-shell microgels with a different initial aspect ratio of the outer frame forming polymeric shell $L_{frame}/l_{frame} = A) 2.8, B)2.5$ and $C) 2.2$. } 
	\label{SI_sim_model}
\end{figure}

\noindent \textit{Microgel description.}
The anisotropic microgels are composed of polymeric shell and a solid core. The beads forming the core and the shell are denoted by $C$ and $S$, respectively. The interaction parameters between the beads of the same species are set to $a_{ii} = 25k_BT/r_c$.
To estimate the interaction parameters between the microgels and the liquids, we first constructed the microgels and then tune the flexibility of the subchains in order to reproduce the swelling/deswelling behaviour of the microgels in the liquids. To model the  swelling / deswelling process we alter the polymer-liquid repulsion ${a_{Si}}$ in a range between $a_{Si} = 25 k_BT/r_c$ and $a_{Si} = 30.2k_BT/r_c$. \\
The anisotropic microgels with elliptical solid core were designed as follows. Similarly to the work Ref.\textsuperscript{S2}, we constructed a unit cell of the diamond crystal lattice where the vertexes correspond to tetrafunctional crosslinkers. Then we assemble two cubic supercells S50 and S25 consisted of 50×50×50 and 25×25×25 unit cells, respectively. Supercell S50 is considered as a template for the solid nanoparticle, while S25 – as a template for the polymeric shell.\\
The dimensions of the hematite-silica core in the experiment were 330$\pm$12nm$\times$75$\pm$8 nm. We created the ellipsoidal nanoparticle having a similar ratio between the long and short semi-axis 4.4$\times$1. The ellipsoidal core was constructed by inscribing the ellipsoid-shaped frame into S50 supercell and cropping all the beads, which are outside of the ellipsoid.  
To construct the polymeric shell around the solid core we used a scaled S25 supercell. All the bonds between the tetrafunctional crosslinker were replaced by the subchains of different lengths. The distribution of subchain lengths is described by a Gaussian distribution with a mean value equal to $N$ and a standard deviation equal to $dN$. Two ellipsoidal frames of different radii were inscribed into the modified supercell. The small frame is necessary for the formation of the void in the polymeric network of the same sizes as the solid nanoparticle ones. The sizes of the large frame control the thickness and the shape of the polymeric shell. All beads, which were located inside the small and outside of the large ellipsoidal frames, were cropped. The rest of the beads form microgel shell. The solid core was inserted into the void of the microgel with further grafting of the dangling chains of the polymeric shell to the nanoparticle surface. The dangling chains of the polymeric shell were physically attached to the nanoparticle surface. Going through all of the free ends of the dangling chains we monitor whether the distance between it and closest bead at the surface of the solid core satisfies the condition: $|r_{end}-r_{surf}|\leq r_{c}$. If so, a bond between the free end of the dangling chains and the bead of the surface of the solid core was formed. Only one bond could be formed between the free end and the solid core.\\
Since the real aspect ratio of the anisotropic microgel is unknown from the experiment, we designed three different microgels with the same shell mass and cross-linking density but different initial aspect ratios of the outer frame $L/d$ = 2.8; 2.5 and 2.2 to capture the peculiarities of the shell dimensions  (Figure \ref{SI_sim_model}). The microgel, constructed using the outer frame with an aspect ratio of 2.8, is featured by a thicker shell at the top of the core compared to its side. The microgel, constructed using the outer frame with an aspect ratio of 2.2 has approximately the same thickness of the shell around the solid particle. The characteristics of the microgels used in simulations are presented in Table~\ref{SI_tab_model}.\\

\begin{table}[H]
     \centering
     \small
    \caption{Types and amount of beads in microgels. The beads forming polymer microgel shell (including cross-linkers) and the solid core are denoted as S and C respectively. }
     \begin{tabular}{ccccc}
       \hline
   & Amount of S beads & Amount of C beads  & \% of cross-linkers \\
        \hline
Microgel CS$_{2.8}$ & 15034 & 8200 & 7.57\% \\
Microgel HS$_{2.8}$ & 15034 & 0 & 7.57\% \\
Microgel CS$_{2.5}$ & 15002 & 8200 & 7.62\% \\
Microgel HS$_{2.5}$ & 15002 & 0 & 7.62\% \\
Microgel CS$_{2.2}$ & 15036 & 8200 & 7.51\% \\
Microgel HS$_{2.2}$ & 15036 & 0 & 7.51\% \\
     \end{tabular}
     \label{SI_tab_model}
     \end{table}

To maintain the bonds between the beads in subchains of the microgels we applied a harmonic bond potential:
\begin{equation}
   U_{ij}^{bonds} = k_{bond~ij}(r_{ij}- r_{eq})^2,
    	\label{simeq5}
\end{equation}
\noindent where $k_{bond~ij} = 35k_BT/r_c^2$, ${i,j \in (S, C)}$ is the bond stiffness, and $r_{eq} = 0.6r_c$ is the equilibrium bond length.

The number of monomers per subchain in real microgels is high enough. To simulate the macroscopic microgels, one has to use coarse-grained descriptions of the subchain.\textsuperscript{S12} Nikolov et al.\textsuperscript{S13} proposed a method to guide the relation between the number of monomers $N$ and the flexibility of the subchains as well as the swelling degree of the microgel. In addition to the harmonic potential a bending angle potential as well as a segmented-repulsive potential (SRP)\textsuperscript{S14} between polymer beads has been introduced:
\begin{equation}
   U_{ijk}^{bend} = k_{bend~ijk}(1 + cos(\Theta)),
    	\label{simeq6}
\end{equation}
\begin{equation}
   U_{ij}^{SRP} = C_{ij}(1 - r_{mid~ij}/r_{SRP}),
    	\label{simeq7}
\end{equation}
\noindent where $k_{bend~ijk}$ is the bending stiffness, and $\Theta$ is the angle between two pairs of connected beads sharing a common bead. ${r_{mid}}$ is the distance between the midpoint of two bonds of connected beads, ${r_{SRP}=0.5r_c}$, the cutoff distance and ${C_{SS} = 100k_BT}$, the strength of the SRP potential between bonds. The SRP potential prevents polymer chains from crossing one another. The bending potential allows controlling the flexibility of the individual polymer subchains by changing the Kunh length. For ${k_{bend} = 0}$, the Kuhn length is relatively short and is approximately equal to 0.6$r_c$. The addition of bending stiffness leads to an increased Kuhn length and swelling degree of the microgel in a swollen state. We found that $N = 6$, $dN = 1$ and ${k_{bend} = 5k_BT}$ are suitable to reproduce both the experimental swelling degree and deformability of the microgels.

The harmonic bond and angle potential had also to be set for the solid core to maintain its shape and volume and prevent its deformation at the interfaces.
\begin{equation}
   U_{ijk}^{angles} = k_{angle~ijk}(\theta- \theta_{eq})^2,
    	\label{simeq8}
\end{equation}
${r_{eq~C} = 0.8r_c; k_{bond~CC} = 100k_BT/r_c^2}$, ${\theta_{eq} = 109.47}$ and ${k_{angle~CCC}  = 100k_BT}$.

Without loss of generality, we set $m = k_{B}T = r_{c} = 1$. Consequently, our unit of time $\tau = r_{c}(m/k_{B}T)^{1/2}$ is equal to 1. The equations of motion are integrated over time with a modified velocity-Verlet algorithm with a time step $\Delta t$ = 0.01 and the values of the noise and damping coefficients were 3.0 and 4.5 respectively.

\textbf{Microgel characterization in bulk}\\
The simulations of single microgels in bulk solution at different $a_{SW}$ values were performed. Larger values of $a_{SW}$ correspond to bad solvents, whereas lower values of $a_{SW}$ correspond to good solvents.  To identify the temperature responsiveness of the microgels at each $a_{SW}$, we calculate hydrodinamic radii and L/d ratios for the microgels having different architecture.\\

\begin{figure}[H]
\centering
    \includegraphics[width=0.9\linewidth, trim={0cm 0cm 0cm 0cm},clip]{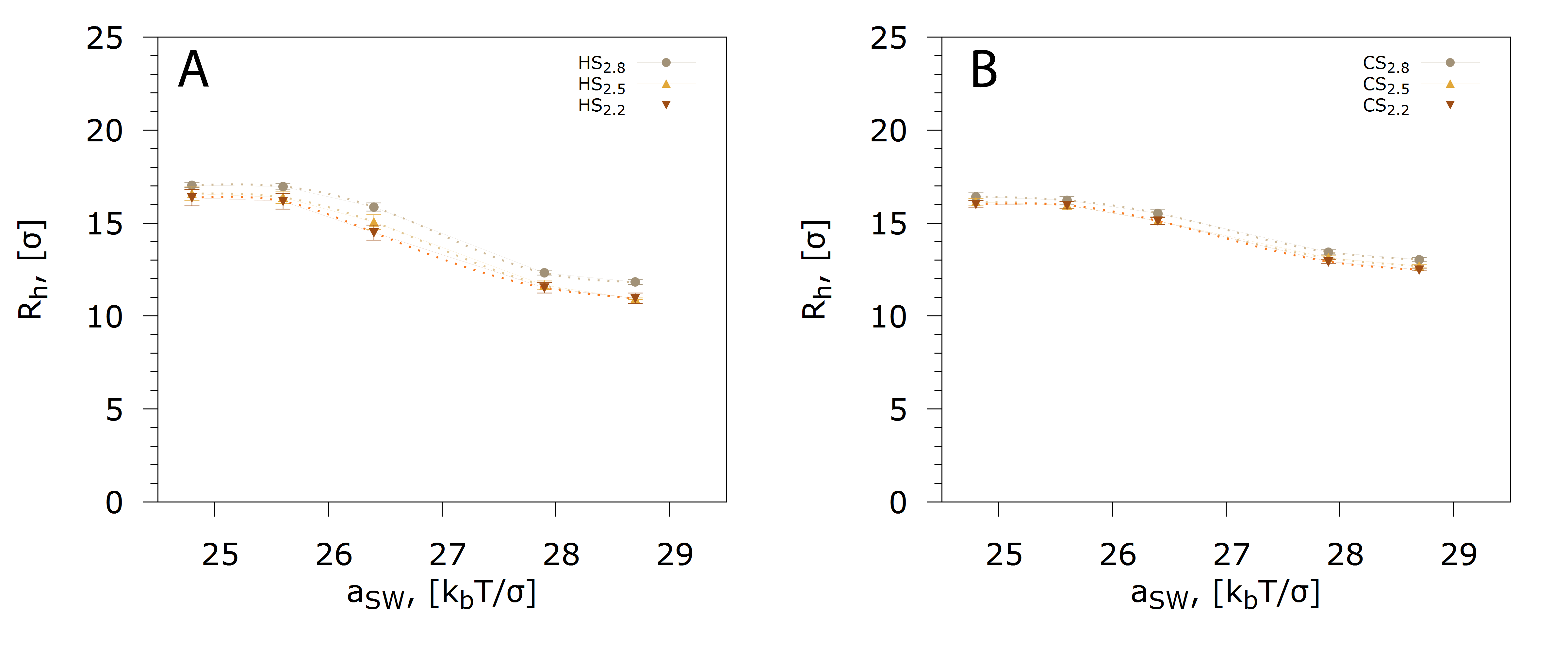}	
	\caption{Hydrodynamic radii of the HS (A) and CS (B) microgels with different initial shell configuration: $L_{frame}/d_{frame} = 2.8, 2.5$ or $2.2$ as a function of interaction parameter $a_{SW}$. }
	\label{SI_sim_radii}
\end{figure}
Figure \ref{SI_sim_radii} shows the combined results for the microgels of interest with and without a solid core. At $a_{SW} < 26.2$ all the microgels are in a swollen state. The radius of the gels follows the relation $C(H)S_{2.8}$>$C(H)S_{2.5}$>$C(H)S_{2.2}$. The continuous decrease in the radius is observed when the $a_{SW} > 26$ and reaches the minimum value at $a_{SW} >27.5$. In experiment the hydrodynamic radius of the $CS-165$ microgel changes from $165\pm3 nm$ in a swollen state to $122\pm1 nm$ in the collapsed state characterised by a swelling degree of 1.4. We choose ${a_{SW}=25.6}$ and $27.9$ to reproduce the microgel properties in a swollen ($T=20°C$) and collapsed ($T=50°C$) states. The swelling degree equal $1.41\pm0.04$, $1.43\pm0.04$ and $1.38\pm0.04$ is obtained for the constructed $CS_{2.8}$,$CS_{2.5}$ and $CS_{2.2}$ microgels respectively.\\
\begin{figure}[H]
\centering
    \includegraphics[width=0.9\linewidth, trim={0cm 0cm 0cm 0cm},clip]{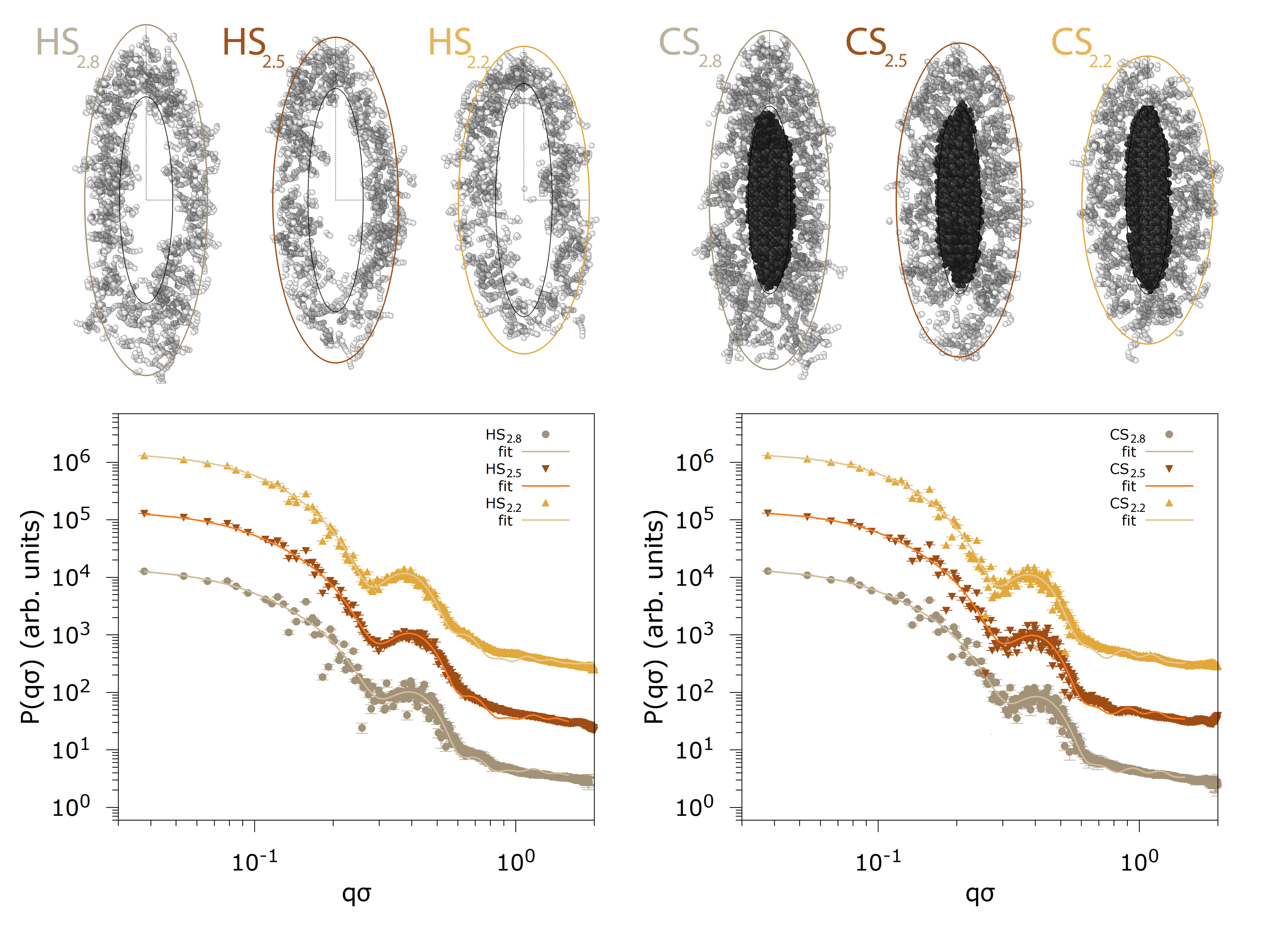}	
	\caption{Equilibrium structure of the swollen HS and CS microgels with different initial shell configuration: $L_{frame}/d_{frame} = 2.8, 2.5$ or $2.2$ and their form factor profiles P(q). The cross-sections of the thickness of $d = 2r_{c}$.  Polymeric shell is depicted in grey, solid nanoparticle – in black, water beads are not shown. $a_{SW} = 25.6$. Symbols are the simulation data, and lines are their fitting curves.}
	\label{SI_sim_2}
\end{figure}
The cross-sections of the equillibrated microgels of the thickness of $d = 2r_{c}$ taken through the center of mass are shown in Figure \ref{SI_sim_2} (swollen state) and \ref{SI_sim_4} (collapsed state).
To estimate the actual radii of the microgel shell, solid particle or the void we have calculated the form factor. First, we started an approximation with a solid core. Fitting the data with the model of core-shell ellipsoids provides the dimensions, ($19.4\pm0.6$)$r_{c}$ for the major and ($4.5\pm0.3$)$r_{c}$ for the minor axes of the solid core. $L_{core}/d_{core} = 4.4$ which is in agreement with the experimental value.\\
In Figure \ref{SI_sim_2} the scattering data for the microgel shell of the CS and HS samples with different initial shell configuration in a swollen state: $L_{frame}/d_{frame} = 2.8, 2.5$ and $2.2$ are presented. We further tried to use the core-shell ellipsoid models to fit it. Nevertheless, in many cases we were not able to find the proper fit to describe the structure correctly. The fuzziness is needed to get into account to follow all details of the scattering points. To overcome it, we first used the LSQR approximation\textsuperscript{S2} of the inner and outer surface of the microgels shell to get the dimensions of the  microgels. Relying on that data, we tried to find the best fit. The bright color solid lines on Figure \ref{SI_sim_2} represent re-fitted form factors. The comparison of the sizes of the major L and minor axes l of the elliptical microgel and the cavity in a swollen state obtained in the simulations is presented in Figure \ref{SI_sim_3}. For the clarity we also plotted the fitted ellipsoids onto the cross-sections of the microgels in Figure \ref{SI_sim_2}.\\
\begin{figure}[H]
\centering
    \includegraphics[width=1\linewidth, trim={0cm 0cm 0cm 0cm},clip]{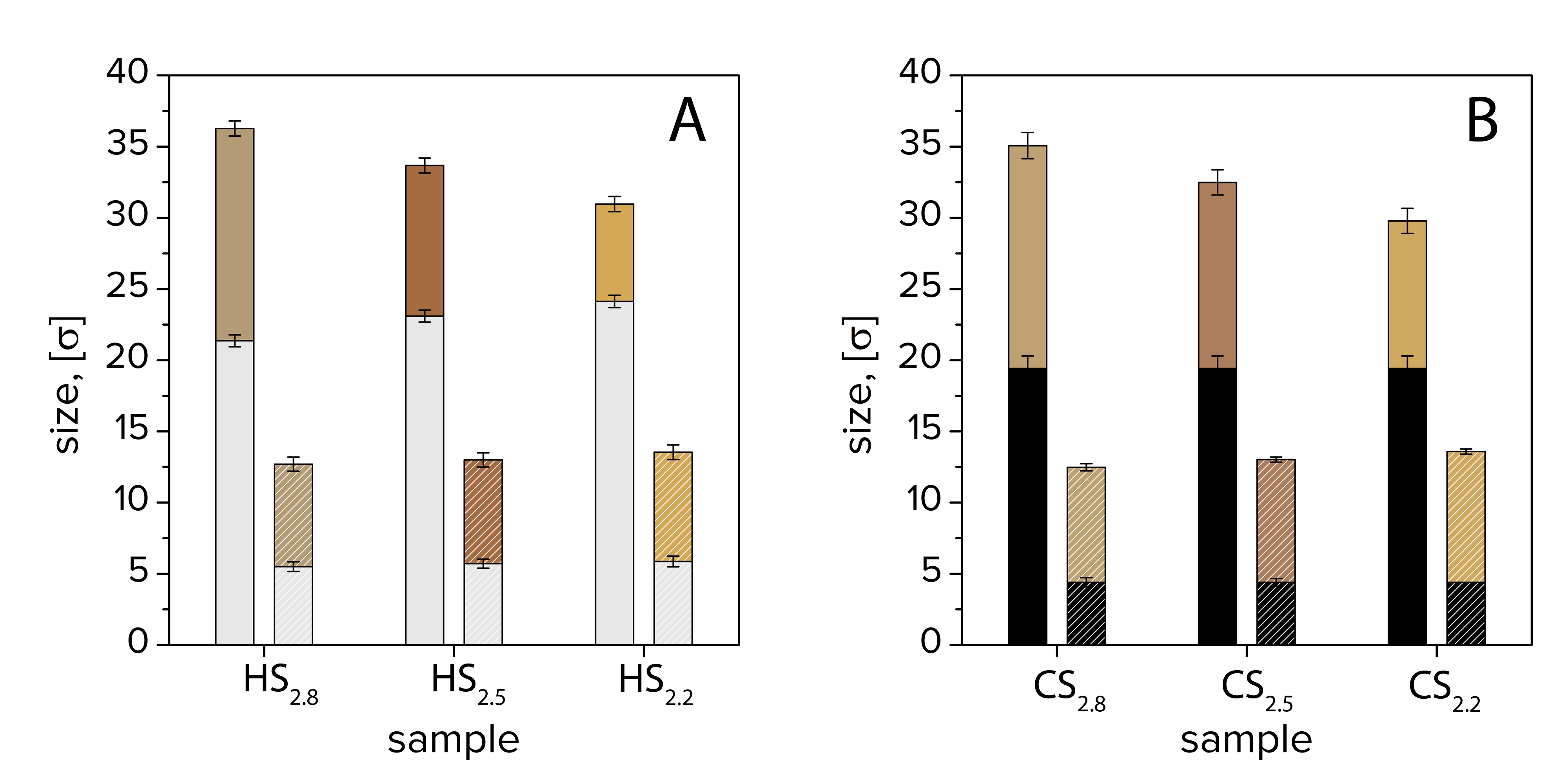}	
	\caption{Sizes of the hollow (A) and the core-shell (B) anisotropic microgels in a swollen state ($a_{SW} = 25.6$) obtained by the fitting of the form factors by means of the model of a spheroid ellipsoid particle with a core shell structure. L and d values of the major and minor axes of the elliptical cavity(gray), solid particles(black) and polymeric shell(colored) are shown as a cumulative bars. Numbers 2.8, 2.5 and 2.2 corresponds to the microgels with different initial shell configuration.}
	\label{SI_sim_3}
\end{figure}

\begin{figure}[H]
\centering
    \includegraphics[width=0.5\linewidth, trim={0cm 0cm 0cm 0cm},clip]{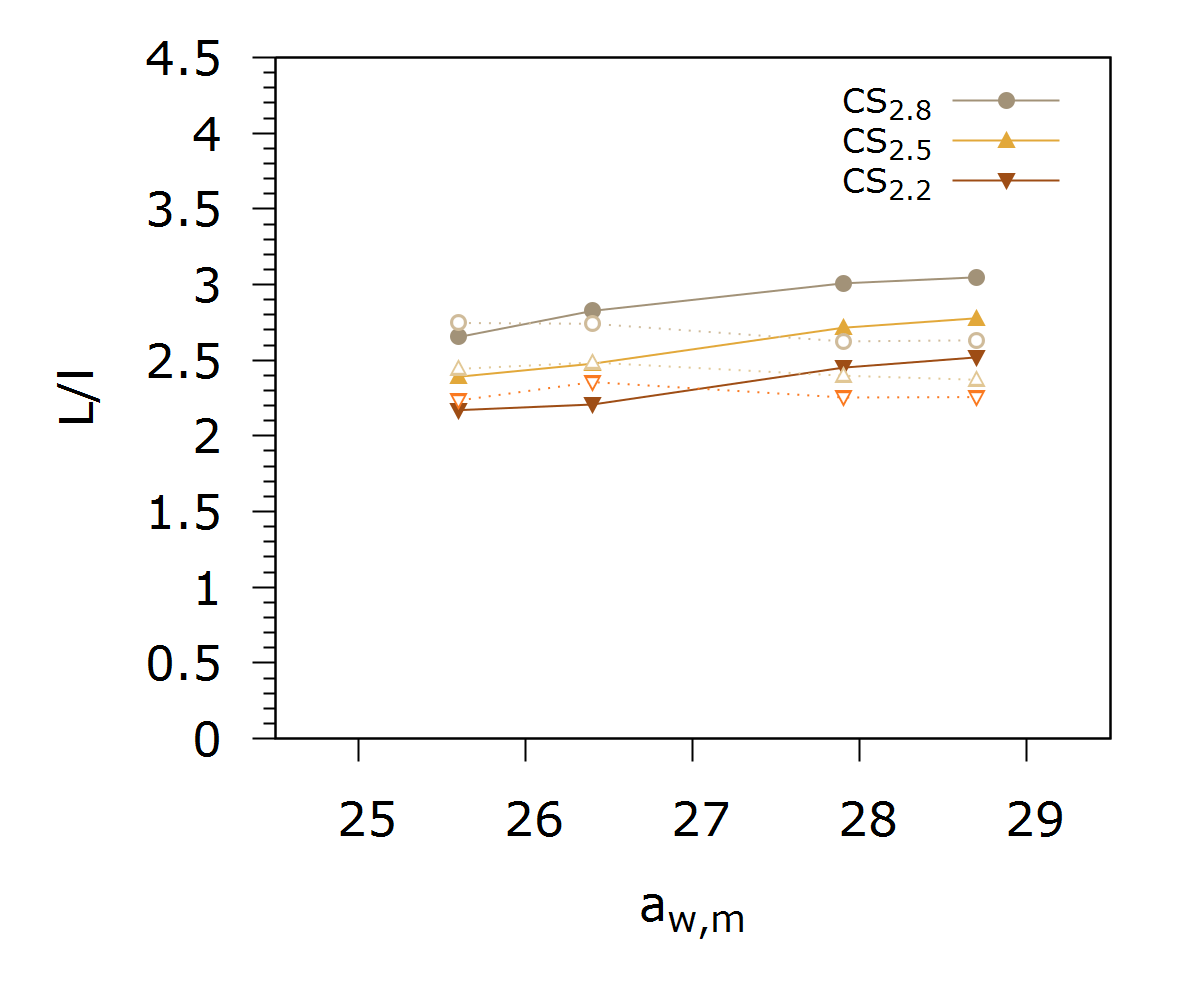}	
	\caption{Shape anisotropy L/d of the hollow (dashed lines) and the core-shell (solid lines) microgels as a function of $a_{SW}$.}
	\label{SI_SI_L-l}
\end{figure}
We have started the quantitative analysis from the results for CS microgels. Obviously, the initial shell configuration determine the distribution of polymers surrounding the ellipsoidal core. After equilibration in a swollen state the ratio between thickness at the tip and the sides, ${\delta}L/{\delta}d$, are equal to 1.93, 1.51 and 1.1 for $CS_{2.8}$, $CS_{2.5}$ and $CS_{2.2}$ samples respectively. The $CS_{2.2}$ can be regarded as a model for the microgel with constant thickness surrounding the ellipsoidal core, while the samples $CS_{2.5}$ and $CS_{2.8}$ – as model represented the microgels having slight and strong shell thickness anisotropy. Increasing the $a_{SW}$ lead to the collapse of the microgel shell accompanied by the increasing of the aspect ratio L/d, Figure \ref{SI_SI_L-l}, in agreement with the results obtained earlier.\textsuperscript{S15} The cross-sections of the equillibrated microgels in a collapse state above the VPTT are shown in Figure \ref{SI_sim_4}. When looking at the cross-sections of the microgels in the collapsed stated, one could find compact dense shell with a sharp surface. This is echoed by the shift of the first minima of the scattering curves to the right and the increase of the slope at high q values.\\
 Independently of the sample etching of the core practically does not change the radii of the shells below VPTT which is in line with the results for the microgels with the thin shell.\textsuperscript{S15} Slight increase of the L/d value was observed for the sample $HS_{2.8}$. Shell inhomogeneities at the tip of the microgel lead to the extra swelling of the shell, Figure \ref{SI_sim_3}. However, more noticeable changes occurred with the size of the cavity. After removal of the core $L_{void}$ > $L_{core}$ for the samples $HS_{2.5}$ and $HS_{2.2}$. Increasing the $a_{SW}$ value leads to the collapse of the hollow microgels. Contrary to the CS microgels, collapse of the HS microgels does not lead to increase of aspect ratio L/d, Figure \ref{SI_SI_L-l}. In other words collapse of the microgel leads to a change in size, yet the anisotropic shape is conserved, confirming the conclusions in ref.\textsuperscript{S15} Moreover cavity was observed at all investigated $a_{SW}$ values. At the same time, the cavity of the microgels in the collapsed sate are much smaller comparing to the solid core. The collapse of the $HS_{2.8}$ leads to the formation of the cavity with a small asphericity value, while collapse of the $HS_{2.2}$ leads to a cavity with narrow elongated shape.

\begin{figure}[H]
\centering
    \includegraphics[width=0.9\linewidth, trim={0cm 0cm 0cm 0cm},clip]{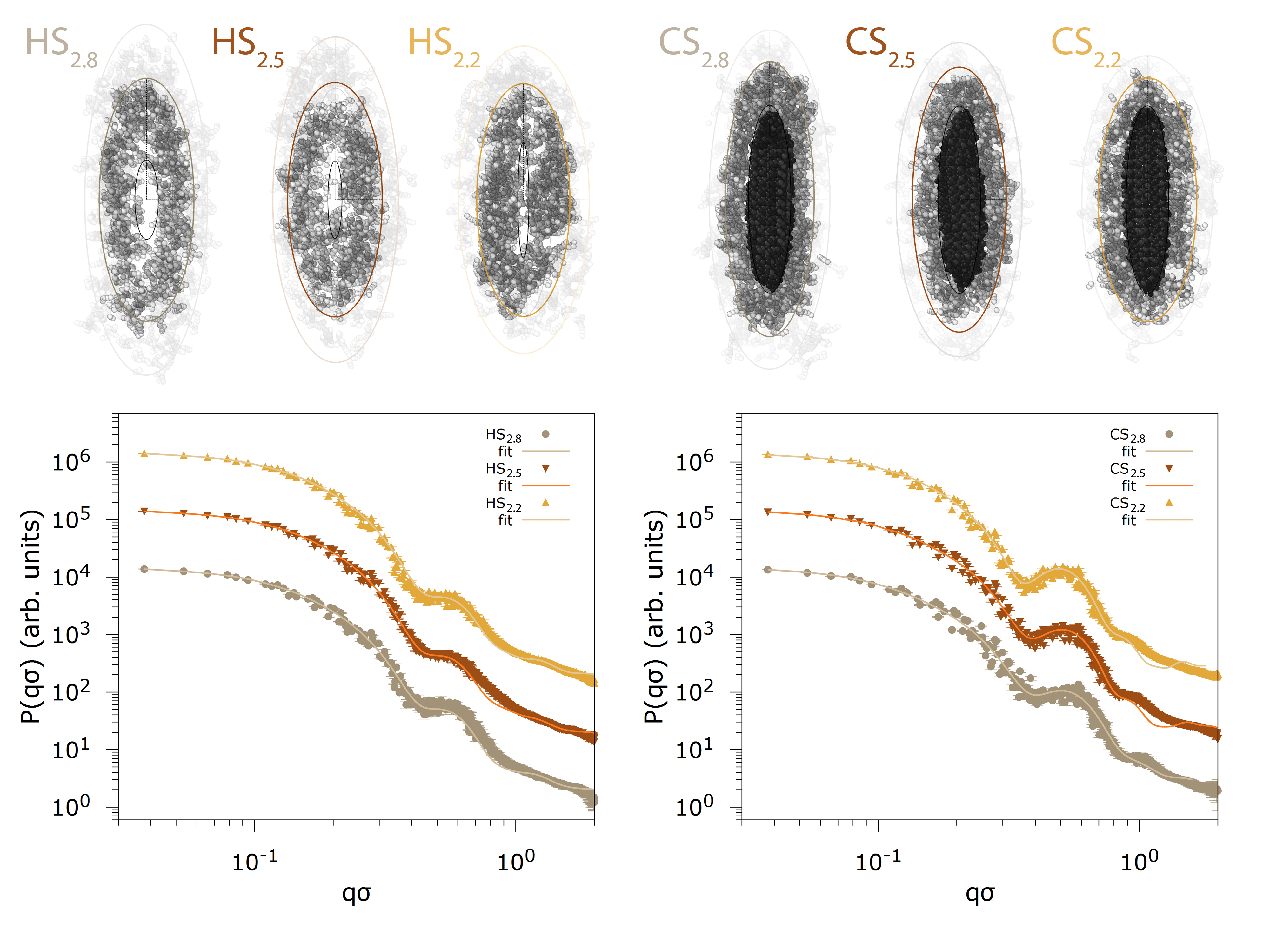}	
	\caption{Equilibrium structure of the swollen HS and CS microgels with different initial shell configuration: $L_{frame}/d_{frame} = 2.8, 2.5$ or $2.2$ and their form factor profiles P(q). The cross-sections of the thickness of $d = 2r_{c}$.  Polymeric shell depicted in grey, solid nanoparticle – in black, water beads are not shown. $a_{SW} = 27.9$. Symbols are the simulation data, and lines are their fitting curves.}
	\label{SI_sim_4}
\end{figure}

\begin{figure}[H]
\centering
    \includegraphics[width=1\linewidth, trim={0cm 0cm 0cm 0cm},clip]{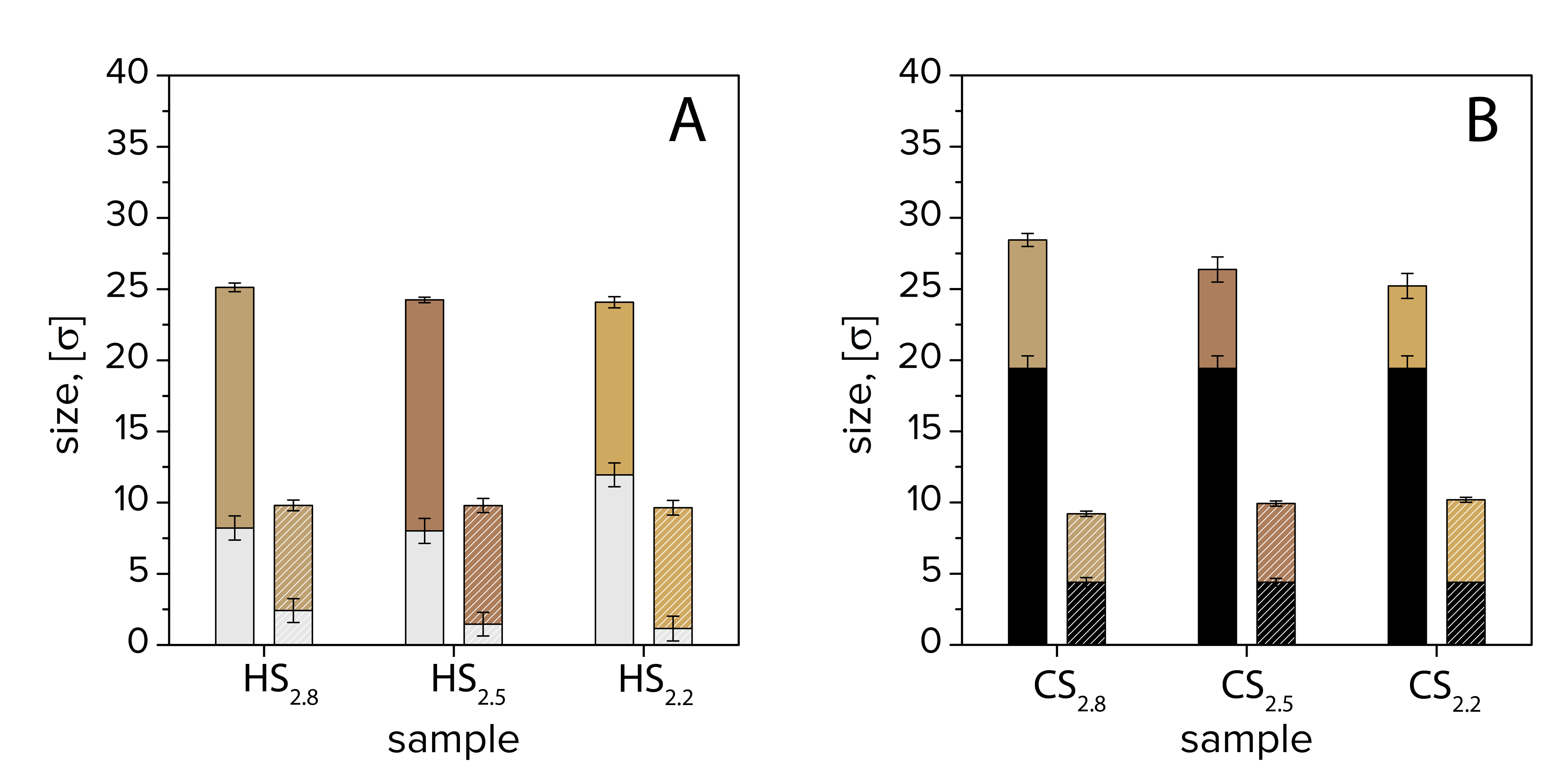}	
	\caption{Sizes of the hollow (A) and the core-shell (B) anisotropic microgels in a collapsed state ($a_{SW} = 27.9$) obtained by the fitting of the form factors by means of the model of a spheroid ellipsoid particle with a core shell structure. L and d values of the major and minor axes of the elliptical cavity(gray), solid particles(black) and polymeric shell(colored) are shown as a cumulative bars. Numbers 2.8, 2.5 and 2.2 corresponds to the microgels with different initial shell configuration.}
	\label{SI_sim_5}
\end{figure}

\textbf{Interface deformation}\\
The simulations of single microgels at the interface were performed. The estimation of the deformation of the water/decane interface were done. To identify the area of deformation, we constructed the surface mesh of the interface using OVITO tool. The procedure for surface construction from a set particles is described in Ref.\textsuperscript{S16} The radius of the probe sphere employed in the surface reconstruction algorithm was selected as 1. The smoothing level was equal to 100. To determine the surface area and plot the color height maps, the ParaView v.5.10.0 were used.\textsuperscript{S17}\\

\begin{figure}[H]
\centering
    \includegraphics[width=0.9\linewidth, trim={0cm 0cm 0cm 0cm},clip]{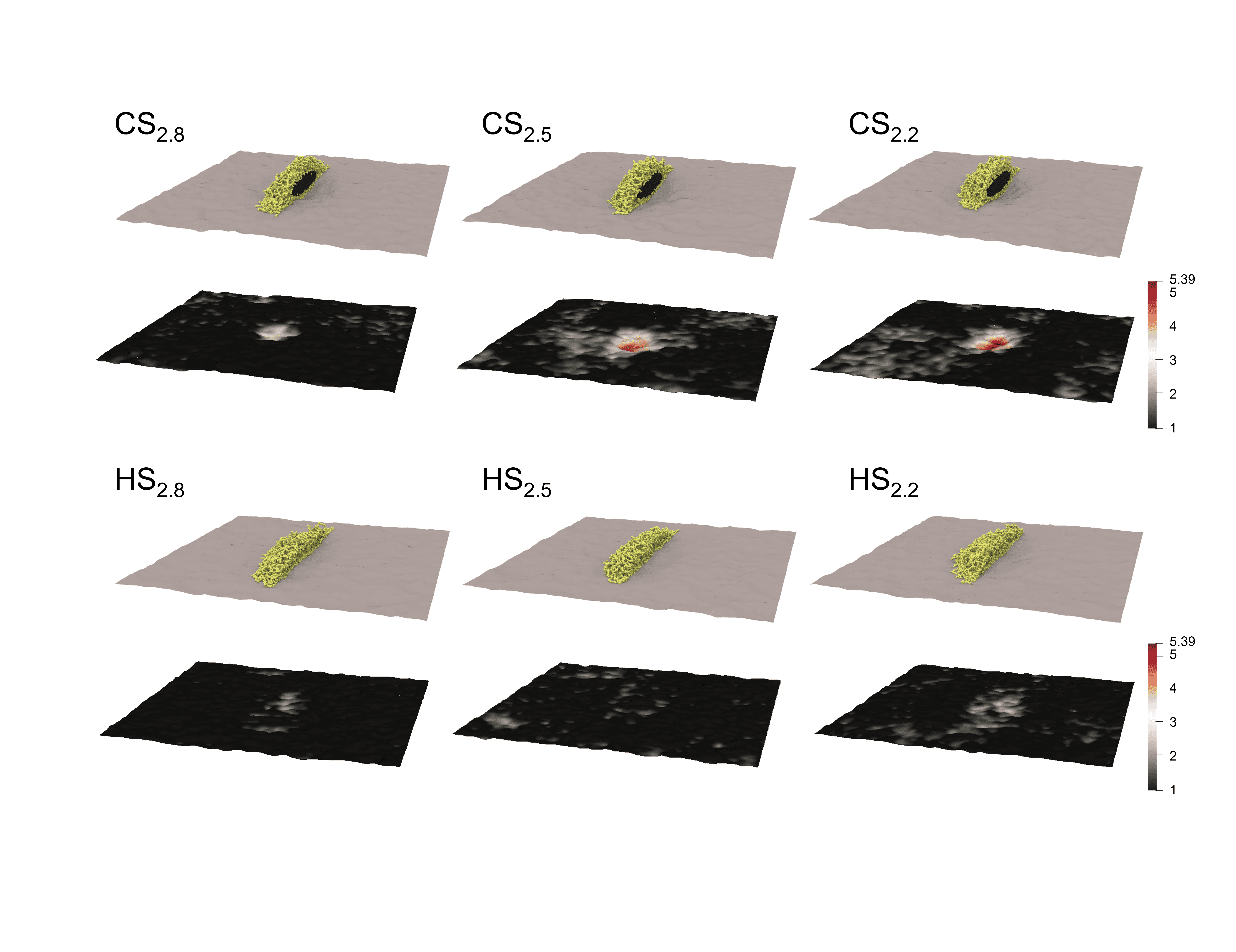}	
	\caption{First and third rows: 3D view of the halves of the CS and HS microgels at the water/decane interface. Second and fourth rows: red and white and black gradient color code height maps of the interfaces. The interfaces are shown as smoothed surface mesh. }
	\label{SI_def1}
\end{figure}

\begin{figure}[H]
\centering
    \includegraphics[width=0.9\linewidth, trim={0cm 0cm 0cm 0cm},clip]{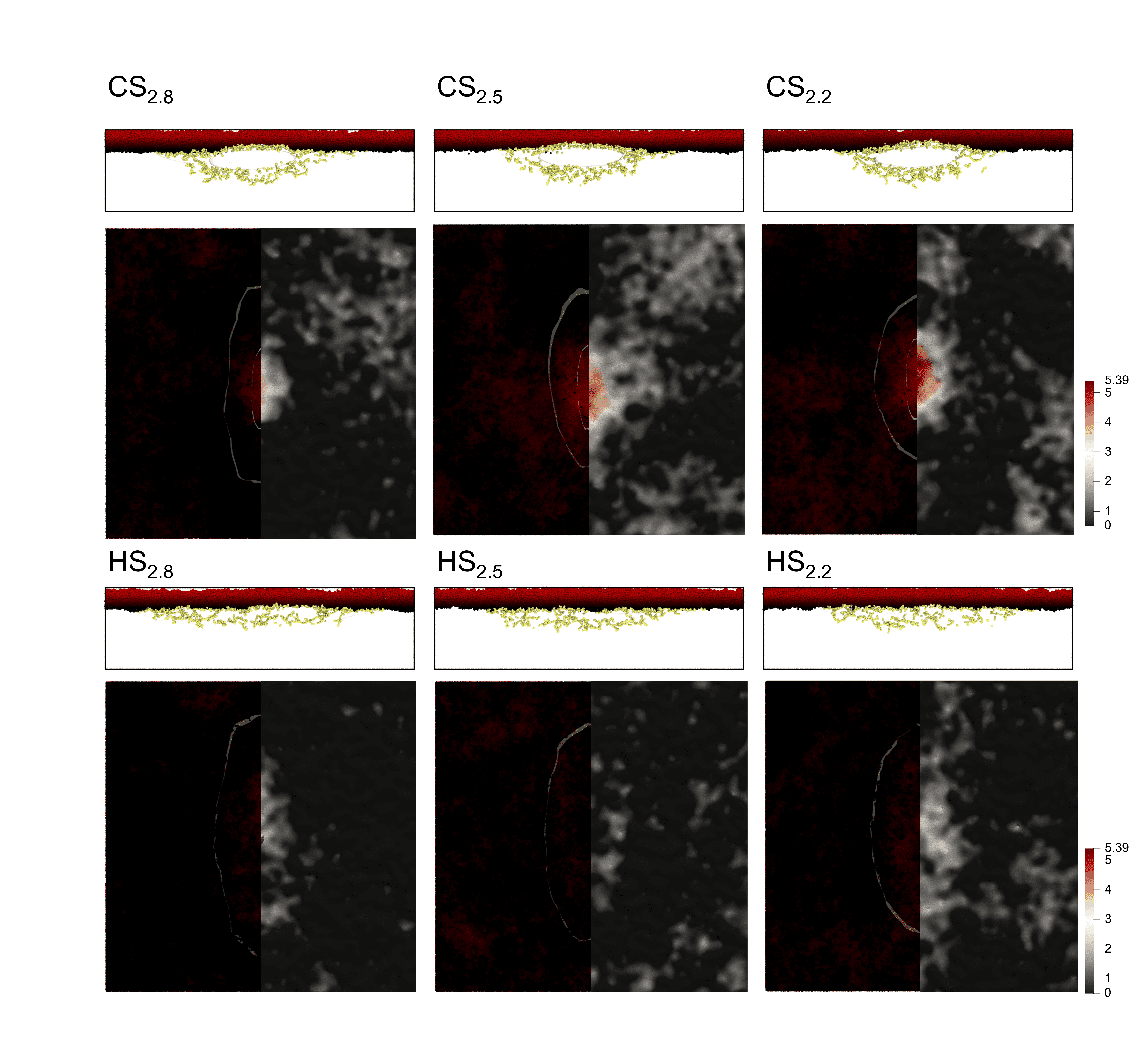}	
	\caption{First and third rows: Longitudinal cross-sections of $CS$ and  $HS$ microgels. Solid particles are not shown. The decane molecules were plotted in red and black gradient color code accordingly to their z position with respect to the interface. Second and fourth rows: red and black gradient color code height maps of the interface. One part of the images is top view to the decane phase on which the outlines of the shell and the solid core are plotted. Another part is the smoothed water/decane surface mesh }
	\label{SI_def2}
\end{figure}

\begin{figure}[H]
\centering
    \includegraphics[width=0.9\linewidth, trim={0cm 0cm 0cm 0cm},clip]{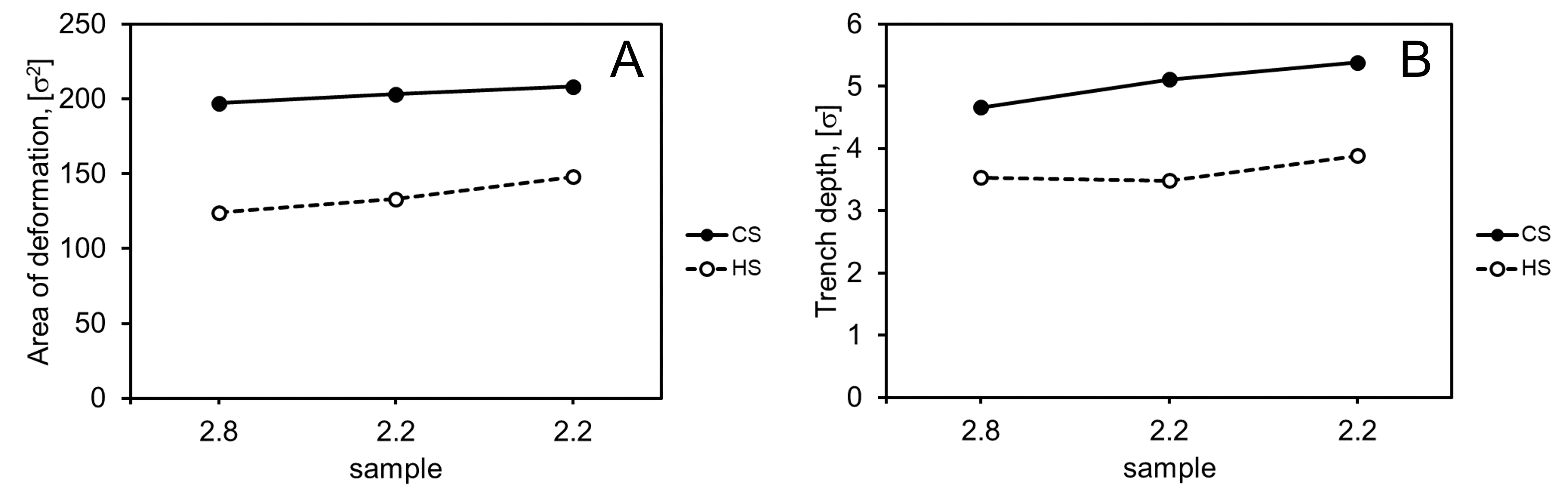}	
	\caption{(A) Area of deformation and (B) trench depth of the interfaces in case of adsorption of $CS$ (full symbols, solid line) and $HS$ anisotropic microgels (hollow symbols, dashed line) for the different designs. }
	\label{SI_def3}
\end{figure}

\noindent{\Large\textbf{References}}

\noindent(S1) Bochenek, S.; Scotti, A.; Ogieglo, W.; Fernández-Rodríguez, M. A.; Schulte, M. F.; Gumerov, R. A.; Bushuev, N. V.; Potemkin, I. I.; Wessling, M.; Isa, L.; Richtering, W. Effect of the 3D Swelling of Microgels on Their 2D Phase Behavior at the Liquid–Liquid Interface. \textit{Langmuir} \textbf{2019}, \textit{35}, 16780–16792.

\noindent(S2) Nickel, A. C.; Kratzenberg, T.; Bochenek, S.; Schmidt, M. M.; Rudov, A. A.; Falkenstein, A.; Potemkin, I. I.; Crassous, J. J.; Richtering, W. Anisotropic Microgels show their Soft Side. \textit{Langmuir} \textbf{2021}, https://doi.org/10.1021/acs.langmuir.1c01748.

\noindent(S3) Geisel, K.; Rudov, A. A.; Potemkin, I. I.; Richtering, W. Hollow and Core–Shell Microgels at Oil–Water Interfaces: Spreading of Soft Particles Reduces the Compressibility of the Monolayer. \textit{Langmuir} \textbf{2015}, \textit{31}, 13145–13154.

\noindent(S4) Espanol, P.; Warren, P. Perspective: Dissipative particle dynamics. \textit{J. Chem. Phys.} \textbf{2017}, \textit{146}, 150901.

\noindent(S5)  Maiti, A.; McGrother, S. Bead-Bead Interaction Parameters in Dissipative Particle Dynamics: Relation to Bead-Size, Solubility Parameter, and Surface Tension. \textit{J. Chem. Phys.} \textbf{2004}, \textit{120}, 1594–1601.

\noindent(S6) Travis, K. P.; Bankhead, M.; Good, K.; Owens, S. L. New parametrization method for dissipative particle dynamics. \textit{J. Chem. Phys.} \textbf{2007}, \textit{127}.

\noindent(S7) Liyana-Arachchi, T. P.; Jamadagni, S. N.; Eike, D.; Koenig, P. H.; Ilja Siepmann, J. Liquid-liquid equilibria for soft-repulsive particles: Improved equation of state and methodology for representing molecules of different sizes and chemistry in dissipative particle dynamics. \textit{J. Chem. Phys.} \textbf{2015}, \textit{142}.

\noindent(S8) Khedr, A.; Striolo, A. DPD Parameters Estimation for Simultaneously Simulating Water-Oil Interfaces and Aqueous Nonionic Surfactants. \textit{J.  Chem. Theory Comput.} \textbf{2018}, \textit{14}, 6460–6471.

\noindent(S9) Groot, R. D.; Warren, P. B. Dissipative Particle Dynamics: Bridging the Gap between Atomistic and Mesoscopic Simulation. \textit{J. Chem. Phys.} \textbf{1997}, \textit{107}, 4423–4435.

\noindent(S10) Zeppieri, S.; Rodríguez, J.; López De Ramos, A. Interfacial tension of alkane + water systems. \textit{Journal of Chemical and Engineering Data} \textbf{2001}, \textit{46}, 1086–1088.

\noindent(S11) Alasiri, H.; Chapman, W. G. Dissipative Particle Dynamics (DPD) Study of the Interfacial Tension for Alkane/Water Systems by Using COSMO-RS to Calculate Interaction Parameters. \textit{J. Mol. Liq.} \textbf{2017}, \textit{246}, 131–139.

\noindent(S12) Spaeth, J.; Dale, T.; Kevrekidis, I.; Panagiotopoulos, A. Coarse-graining of chain models in dissipative particle dynamics simulations.  \textit{Ind. Eng. Chem. Res.} \textbf{2011}, \textit{50}, 69-77.

\noindent(S13) Nikolov, S.; Fernandez-Nieves, A.; Alexeev, A. Mesoscale modeling of microgel mechanics and kinetics through the swelling transition. \textit{Appl. Math. Mech. (Engl. Ed.)} \textbf{2018}, \textit{39},  47–62.

\noindent(S14) Sirk, T.; Sliozberg, Y.; Brennan, J.; Lisal, M.; Andzelm, J. An enhanced entangled polymer model for dissipative particle dynamics. \textit{J. Chem. Phys.} \textbf{2012}, \textit{136},  134903.

\noindent(S15) Nickel, A. C.; Scotti, A.; Houston, J. E.; Ito, T.; Crassous, J.; Pedersen, J. S.; Richtering, W. Anisotropic Hollow Microgels That Can Adapt Their Size, Shape, and Softness. \textit{Nano Lett.} \textbf{2019}, \textit{19}, 8161–8170.

\noindent(S16) Stukowski, A. Computational Analysis Methods in Atomistic Modeling of Crystals. \textit{JOM} \textbf{2014}, \textit{66}, 399–407.

\noindent(S17) Ahrens, J.; Geveci, B.; Law, C. \textit{ParaView: An End-User Tool for Large Data Visualization, Visualization Handbook}; Elsevier, 2005.
